\documentclass[a4paper,fleqn,usenatbib]{mnras}

\usepackage{newtxtext,newtxmath}

\usepackage[T1]{fontenc}
\usepackage{ae,aecompl}

\usepackage{graphicx}	
\usepackage{amsmath}	
\usepackage{amssymb}	

\usepackage{color}
\usepackage{ulem}
\usepackage{multicol}
\usepackage{caption}

\include{epsfig.sty}

\title[Errors, chaos and the collisionless limit]{Errors, chaos and the collisionless limit}

\author[Amr A. El-Zant, Mark J. Everritt and Summer M. Kassem]
 {Amr A. El-Zant$^{1}$\thanks{E-mail: amr.elzant@bue.edu.eg}, Mark J. Everitt$^{2}$ and  Summer M. Kassem$^{3}$
\\
$^{1}$Centre for Theoretical Physics, The British University in Egypt, Sherouk City 11837, Cairo, Egypt\\
$^{2}$Department of Physics, Loughborough University, Loughborough, Leicestershire LE11 3TU, United Kingdom\\
$^{3}$Physics Department, The American University in Cairo, New Cairo 11835, Cairo, Egypt 
}

\date{Accepted XXX. Received YYY; in original form ZZZ}

\pubyear{2016}

\begin{document}
\label{firstpage}
\pagerange{\pageref{firstpage}--\pageref{lastpage}}

\maketitle

\begin{abstract}

We simultaneously study the  dynamics of the growth of errors and the question of the
faithfulness of simulations of $N$-body systems.  The errors are quantified through the numerical 
reversibility of small-$N$ spherical systems, 
 and by comparing fixed-timestep runs with different stepsizes. 
The errors add randomly, before exponential divergence sets in,    
with exponentiation rate virtually independent of $N$,  but scale saturating as $\sim 1/\sqrt{N}$, 
in line with theoretical estimates presented.
In a third phase, the growth rate is initially driven by 
multiplicative enhancement of errors, as in the exponential stage. It is then 
qualitatively different for the phase space variables and 
mean field conserved quantities (energy and momentum); for the former, 
the errors grow systematically through phase mixing,  
for the latter they grow diffusively. For energy, the $N$-variation of the `relaxation time' of error growth follows 
the $N$-scaling of two-body relaxation.  This is also true for angular momentum 
in the fixed stepsize runs, although the associated error threshold is higher and the relaxation time smaller.
Due to shrinking saturation scales,  the information loss associated 
with the exponential instability decreases with $N$ and the 
dynamical entropy vanishes at any finite resolution as $N \rightarrow \infty$. 
 A distribution function depending 
on the integrals of motion in the smooth potential is decreasingly affected. 
In this sense there is convergence to the collisionless limit, 
despite the persistence of  exponential instability on infinitesimal scales. 
Nevertheless, the slow $N$-variation in its saturation points to the slowness of the convergence.

\end{abstract}





\section{Introduction}

 The dynamics of the growth of small perturbations to the trajectories of 
Newtonian $N$-body gravitational systems is of importance, both 
from a theoretical point of view and in terms of its implications on the accuracy 
and faithfulness of numerical simulations. Theoretically, $N$-body systems are expected 
to tend to the collisionless limit as $N$ increases, yet this can be rigorously proven only 
when the two body potential is smooth (Braun \& Hepp 1977),
which is the case in softened systems but not for point particle interactions.  
For unsoftened systems, which is the case of actual stellar systems for example,  
a peculiar phenomenon concerns the persistence of the exponential divergence between 
nearby trajectories. This takes place at rate that does not decrease with $N$, even for configurations 
with corresponding smooth potentials that support only 
regular orbits (Miller 1964; Kandrup \& Smith 1991; Goodman, Heggie \& Hut 1993, hereafter GHH;
Hemsendorf \& Merritt 2002). It thus appears to be in direct contradiction 
with the predictions of the collisionless limit; in this limit,  the orbits are characteristic curves 
of the collisionless Boltzmann equation and, being regular, should not display exponential 
instability, which is characteristic of chaotic systems. 

Numerically identified dynamical 
chaos normally comes with important physical consequences (e.g., Lichtenberg \& Lieberman 1992; Contopoulos 2002).
In principle, therefore, it is not {\it a priori} obvious  that 
the apparently 'chaotic' nature of the $N$-body problem does not lead to any physically relevant 
phenomena.  On the other hand, the effects associated with the divergence of  trajectories
can appear as artifacts that affect particle simulations, where particle numbers are usually
much smaller than in the physical systems modelled, and various approximations, effectively  perturbing the dynamical system 
in a complicated manner, are made. Thus, the collisionless limit may not be reproduced in simulations 
because the dynamics do not converge to  it, even in principle, or because the numerical image does not.

$N$-body simulations have long been  
indispensable to understanding structure formation in the universe, and the evolution of galaxies, galaxy clusters 
and stellar clusters (e.g. Heggie \& Hut 2003; Dehnen \& Read 2011; Springel 2016). 
 Yet, despite decades of  steady increase in sophistication, relatively little 
progress has been made in assessing their accuracy, or even the qualitative faithfulness of the results obtained to the `true' 
dynamics.   Indeed, more efficient force calculation schemes normally entail larger 
errors than direct force evaluations  (see however Dehnen 2014 for a counter-example). 
In addition,  an estimate of the local or global 
error in the integration of the equations of motion is not in general available.  

The exponential instability places a fundamental limitations
on the accuracy of simulations  by inflating any errors arising 
from approximate force evaluations, as well roundoff errors arising from finite arithmetic and 
truncation errors associated with finite integration timesteps. 
In the absence of any intrinsic divergence in the solutions,
the errors scale linearly, or slower, with the number of timesteps (e.g., Press et. al. 2007).
This will also be the case with the action and angle variables of systems supporting 
regular trajectories (for a general discussion of regular versus chaotic motion in 
the context of stellar dynamics see, e.g., Binney \& Tremaine 2008; hereafter BT). 
As we discuss in Section~\ref{sec:sim_estimates}, this implies that accurate solutions can 
be obtained by increasing the order of the integrator and decreasing the time-step. 
A remedy that does not apply in the presence of the exponential instability.

Despite this, it is widely believed 
that $N$-body simulations do reproduce the correct qualitative behaviour of the modelled systems. 
In some sense, this is a statistical statement. It embodies the belief that even when the correct trajectories 
cannot be captured, statistical faithfulness may be retained in terms of the proper rendering of the gross structure 
of the system, as characterized by the distribution function and the moments derived from it. 
In this context, confidence in the faithfulness of the simulations stems from the 
general stability of gross characteristics (e.g., density and mean velocities and velocity dispersions), 
when large-$N$ systems are simulated from different initial conditions taken from the same statistical distributions. 
This stability persists even when different simulation techniques are used,  though there is  no guarantee that all results are not equally wrong.  
More rigorous justification of the faithfulness of results may be sought in terms of the existence of shadowing trajectories.  These 
are exact solutions that start from nearby initial conditions and remain close  to the numerical solutions for the
timescales of interest.  Their presence, however, is  generally quite difficult to prove and has only been studied in idealized cases (Quinlan \& Tremaine 1992; Hayes 2003). 
On the other hand, it is known that  at least in models where collective instabilities are present, numerical implementation 
and associated exponentially inflated error can play an important role in determining the macroscopic evolution  
(Sellwood \& Debattista 2009; Benhaiem et al. 2018).  Understanding this role may also be of importance in the context 
of interpreting the results of cosmological simulations  (Thi{\'e}baut  et. al. 2008;  Keller et. al. 2018;  Genel et. al. 2018).

The robustness of the results of simulations can also be viewed in terms of the orbital structure, 
which links intrinsic dynamical properties, related to the stability of trajectories, 
to questions regarding the accuracy of numerical representation.
In general, the phase space of Hamiltonian systems can exhibit a variety of regular and chaotic orbital families (e.g., Lichtenberg \& Lieberman 1992; BT), and a numerical method can be said to faithfully represent the dynamics 
if it qualitatively captures these structures. 
A $N$-body trajectory formally moves in a $6 N$ dimensional 
phase space; and is, most generally, only constrained by globally conserved quantities, 
such as the total energy and angular momentum.   Nevertheless, in the collisionless limit, as 
$N \rightarrow \infty$, the $N$-body problem should effectively reduce to that of $N$ independent
systems, as each particle moves in the (self consistent) mean field produced by all 
particles. The relevant phase space structure is six dimensional.  
If, furthermore, this mean field is time independent  and possesses sufficient symmetry, the 
system is `separable' and the problem is further reduced to that of $3 N$ independent one dimensional oscillations or rotations.  
In a six dimensional phase space,  particle trajectories are then confined on tori that can be characterised 
by conserved quantities, and parametrised for example by action variables, reflecting the amplitudes
of the oscillations, and angles following the phases.   
The difference between phases  of orbits parametrised by different values
of the actions diverge only linearly in time (BT).   This situation, though restricted, forms the basis of much 
of classical galactic dynamics.  It also relates the two aspects of our study: such a configuration can be said to approach the collisionless limit if its dynamics can be parametrised in terms of quantities conserved along trajectories in the smoothed (mean field) potential, and
a numerical simulation of such a system is 'faithful' if it captures this structure. 
A particularly simple and important example is that of a large-$N$ spherical system 
in a steady state, where each trajectory should conserve energy and angular momentum. 
These mean field conserved quantities are the integrals of motion of the smooth potential.
This is the case primarily considered here. 
   
A key feature suggesting that the effects of the exponential instability do decrease in importance 
with increasing $N$, despite the non-saturating rate, 
comes from theoretical arguments and idealised numerical experiments  suggesting 
that its spatial scale tends to saturate as the particle number increases, even if its rate does not  
(GHH; Valluri \& Merritt 2000; Kandrup \& Sideris 2001; Sideris \& Kandrup 2002). 
Indeed, trajectories moving in a `gravitational Lorentz gas' of fixed particles appear to increasingly follow their smooth-potential counterparts  as the number of fixed particles is increased. 
The exponential instability timescale itself also increases with $N$ when 
softening is introduced (GHH; Huang, Dubinski \& Carlberg 1993; El-Zant 2002).

Thus, although divergence between nearby trajectories implies loss of detailed 
information regarding these, 
as we will see here, 
this loss decreases with $N$ as the instability saturates 
on progressively smaller scales.  Furthermore, 
we find that errors in mean field conserved quantities 
eventually follow a slow diffusive growth 
(on a timescale comparable to the two body relaxation time in case of particle energies).  
This is the case even if the initial exponential divergence affects these 
quantities in the same way as the phase space variables.
 Thus, from a dynamical systems point of view, the orbital structure  can be  
well preserved, and the information loss limited, despite the persistence of the local exponential instability on the infinitesimal scale. 
From a statistical viewpoint,  if the integrals of motion are well conserved, 
the basic features of the collisionless description should hold, as in 
this limit spherical steady state systems are described 
by a distribution function depending on 
those quantities.  From a numerical perspective, 
if the integration preserves the mean field 
conserved quantities in a way that  
permits the reproduction of the statistical properties described by the distribution function, 
the representation is faithful to the collisionless description in this specific sense.

In this context, the purpose of the present study was twofold: to examine 
the dynamical divergence in both the phase space variables and the mean field conserved quantities
of initially nearby trajectories, and to do this in a way that relates the associated information loss to 
 the accuracy and faithfulness of the numerical representation of the dynamics.  
For this purpose, we examined the time-reversibility of a large number of spherical 
$N$-body systems integrated with high precision using an adaptive Runge-Kutta method 
with present error tolerance. 
In principle, these systems
should be exactly time reversible when 
the velocities are reversed. Nevertheless, this is not generally the case with their numerical image;   
the exponential divergence of trajectories, and associated loss of information,
renders them irreversible if a general integrator is used (e.g. Hoover \& Hoover 2012).    
In the context thus set, 
a simulation can be considered `inaccurate' on timescales on which information loss leads to irreversibility in the phase space variables, 
but is only `unfaithful' over timescales on which the mean field conserved quantities in the reversed system differ considerably from the originals.  
We use this diagnostic in drawing conclusions concerning the implications, physical and numerical, of `chaos' 
in $N$-body gravitational systems.  In addition, in order to better relate to contemporary $N$-body simulations
of collisionless systems, we also examine the same aforementioned issues using the widely employed 
symplectic leapfrog integrator (which has reversible truncation errors when applied to reversible dynamical systems).  
This is done by comparing the evolution of 
systems that are started from the same initial conditions but evolved forward in time with different stepsize. 
We also compare the results using the leapfrog integrator with those obtained using second and fourth order 
Runge-Kutta methods with fixed stepsize.  

In the next section we describe the numerical setup for our simulations and the method through which the estimates of 
error growth are evaluated. Section~\ref{sec:main} constitutes the bulk of this study. We start by delineating the differences, 
crucial to the interpretation of our results, between systematic, diffusive and exponential evolution of the growth of 
errors, presenting  estimates of the expected numerical errors. We then report the numerical results, which 
are interpreted by adapting a model due to GHH, which is shown to predict the persistence of 
systematic growth after the saturation of the exponential divergence. In the case of the phase space variables,  
this is followed by phase mixing; for the mean field conserved quantities it is followed by the 
onset of diffusion. We then evaluate `relaxation times', associated with the growth of errors in the different quantities, 
contrasting the growth of velocity errors
with the that in angular momentum and energy.  The results of the fixed stepsize runs
are presented, before summarising our findings in Section~\ref{sec:sum_disc}. 
In Section~\ref{sec:chao_coll}  we  discuss whether $N$-body systems  can be characterised as `chaotic' on any 
non-infinitesimal scale on the phase space 
as $N \rightarrow \infty$,  given the progressively smaller loss of information, particularly as 
associated with the mean field conserved quantities (a more formal discussion of the relation between chaos,  
information loss and effective irreversibility is given in the Appendix).

\section{Numerical setup}
\subsection{Reversibility as a probe of numerical accuracy and faithfulness}

 Newtonian $N$-body systems obey equations where the time variable appears   
exclusively in second order form. These systems are thus 
time reversible in principle:  for every solution 
evolved up to a time $t_f$, there is a solution --- obtained 
by reversing the velocities at $t_f$  and evolving the system 
for another timespan $t_f$ ---  such that,  for $t  \ge t_f$,
${\bf r} (t) = {\bf r} (2 t_f - t)$ and ${ \dot{\bf r}} (t) = -{\dot{\bf r}} (2 t_f - t)$, 
where ${\bf r}$ is the $3 N$ dimensional coordinate vector of particle positions.

Numerically, however, the situation is generally different. Even assuming that the forces are evaluated to machine precision, 
there will be roundoff errors
in these, as well as in the phase space variables themselves (including the initial conditions). 
Finite time-stepping also leads to truncation errors. 
If the errors are sufficiently inflated by intrinsic divergence of trajectories, 
numerical irreversibility will result, as the accumulating errors in the forward and reversed trajectories diverge.  
It is  of course possible 
to employ  time symmetric algorithms, which are 
formally reversible when applied to time reversible dynamical systems 
(Hernandez \& Bertschinger 2018; Hairer,  Lubich  \&  Wanner 2006, section V.1); 
the symplectic mapping embodied in the widely used leapfrog integrator 
is one such scheme.  Although the truncation errors are then reversible,  floating point 
errors remain, unless integer or fixed point arithmetic is used
(e.g., Earn 1994). Formal reversibility is also generally lost when a variable timestep is introduced, and  devising   
a reversible scheme with variable 
time-stepping is a highly non-trivial task (Dehnen 2017; Hernandez \& Bertschinger 2018).  
This makes it difficult, if not impossible, to estimate local truncation errors using conventional methods, which rely on step-size variation, and at the same time maintain numerical reversibility.

To examine the reversibility of simulated $N$-body systems, 
we use a conventional fourth-fifth order Runge-Kutta integrator with an adaptive timestep and 
 a preset tolerance (Press  et. al.  2007).   
The error estimate at each step is determined by decreasing 
the  stepsize until the tolerance criterion is achieved; namely when the maximum  
error in any one of the phase 
space variables $y$ (any of the position or velocity components) 
is smaller than  $Tol \times (|y| + \epsilon)$, 
with $Tol = 10^{-8}$ and $\epsilon = 10^{-22}$.  This essentially measures the relative 
error in $y$, with provision for situations when its absolute value is very small.
This local estimate may be inserted into theoretical 
expectations of error propagation, which can then be compared 
with error estimates resulting from subtraction of the phase space variables and mean field conserved 
quantities of the forward and reversed trajectories.  
Although the use of a non-geometric integrator with adaptive stepsize can lead to secular drift 
in the total energy, this was found to be minimal. Fig.~\ref{fig:toten} shows examples of the total energy; the energy is conserved 
to better than one part in $10^{-6}$  for $N = 128$ and $10^{-8}$ for $N = 8192$
(the results shown are  for single forward in time runs and not averages over all the 
runs conducted with same $N$, as described below).

To connect to current simulations of collisionless systems, where symplectic integrators are normally used, we 
also integrate pairs of systems started from the same initial conditions and run forward in time with different stepsize. 

\begin{center}
\centering			                    
	\includegraphics[width=1.\columnwidth, trim = 0cm 0.cm 0cm 0cm, clip, angle = 0]{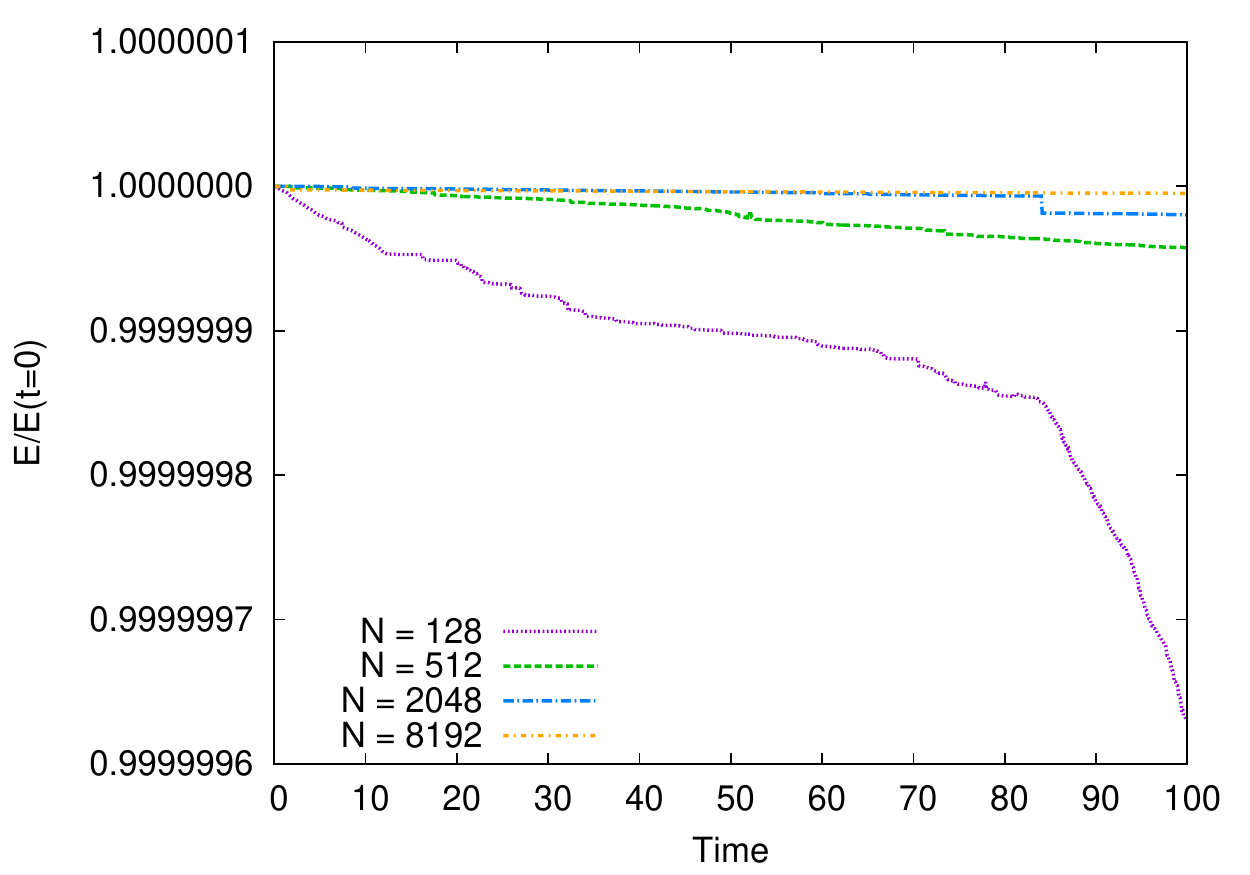}
						\captionof{figure}{Evolution of the total energy in some sample runs, normalised to the initial 
starting energy. The simulations use an adaptive Runge-Kutta routine with preset tolerance  of $10^{-8}$ (relative error) per timestep
and no force softening.} 
    \label{fig:toten}
\end{center}

When a predetermined tolerance is used, we keep the forces unsoftened,  error growth due to close encounters being controlled by varying the adaptive stepsize. This is in line with our goal of examining the 
growth and saturation of  errors in unsoftened systems, which is expected to be  
different from that in softened ones, where the exponentiation rate decreases with $N$ (El-Zant 2002; GHH). 
When a fixed timestep is used we add a small Plummer softening.

\subsection{Initial conditions}
\label{sec:ICS}

We start with random realisations of Plummer spheres (e.g., BT). These are simple models that have 
long formed a benchmark of sorts for theoretical studies of the Newtonian $N$-body problem. 
The mass distribution as a function of radius $r$ is given by 
\begin{equation}
M (r)  = M \frac{r^3}{(r^2+ a^2)^{3/2}},
\label{eq:Plum}
\end{equation}
where $M$ is the total mass of the system and $a$ is a length scale within which the density
distribution varies slowly. We take $G = M = a = 1$. The particle mass is therefore $m = 1/N$. 
We define a mean dynamical time as 
$t_D = 1/\sqrt{G \bar{\rho}(< r)}$, so that each time unit in our simulations corresponds to about $\frac{1}{3.4} 
t_D (r=a)$.  
The virial radius  is at $R_v = \frac{16}{3 \pi}$. The dynamical
time there is $t_D (R_v) = 6.4$ time units,  while the dynamical time at the 
half mass radius $R_h = 1.3 a$ is about 4.4 time units.  The 
standard two body relaxation 
time, $t_r \approx 0.1 \frac{N}{\ln \gamma N} t_D$, ranges from about $31$ time units for the smallest simulated systems
($N = 128$) to $771$ time units for the largest ones ($N = 8192$), if we take $\gamma = 0.11$, as suggested by
Giersz \& Heggie (1994), and $t_D = t_D (R_v)$.  These timescales correspond to about a third and eight times
the timescale of our runs in each direction (forward and reversed).  
The late-time (diffusive) growth of particle energy errors in our simulations, 
which can be used to define an analogous `relaxation time',  is found to scale in a similar manner 
with $N$, but with a timescale that is larger by a factor of 2.6 (cf. figure~\ref{fig:Thresh} 
and discussion in Section~\ref{sec:sum_disc}).


The 
initial conditions are obtained through the technique of Aarseth, Henon \& Wielen (1974). 
The Plummer spheres are truncated at ten core radii, where the distribution becomes poorly sampled. 
This  leads to small systematic initial departures from equilibrium (particularly clear from the oscillations in 
the derivatives near the end of the timescale shown in Fig.~\ref{fig:logderis}).
The Newtonian equations of motion are then integrated, with double precision 
arithmetic and  direct force evaluation. 
Direct force calculation and high accuracy requirements naturally limit the number 
of particles in the systems we probe. In partial compensation for this, we 
improve the statistics by running a number of random realizations for each $N$. This enables 
one to get better statistics with only linear increase (with $N$) in CPU-time, while  
the  time required for the exact force calculations 
scales as $N^2$. We performed $101$ runs for systems with $N=128,\ \ 256, \ \ 512$ and $1024$, $28$ runs
with $N= 2048$, $17$ runs with $N= 4096$ and three runs with $N= 8192$.
Unless otherwise implied, namely in connection with figures~1, 11 and~12, all numerical results shown correspond to averages over all runs.

Each random realization is integrated forward in time for a hundred time units.  When the reversibility 
criterion is used,  velocities are then 
reversed and the system is integrated for another hundred time units.
The errors are then evaluated as discussed below.

\subsection{Error estimates from simulations}

 When the reversibility criterion is used we estimate the errors from the difference in particle positions between the forward and reversed 
trajectories by calculating
\begin{equation}
\xi_x^2 = \frac {\sum_{N}  |{\bf x}_f - {\bf x}_b|^2}{\sum_N |{\bf x}_b|^2}, 
\label{eq:relcoor}
\end{equation}
where ${\bf x}_f$  corresponds to a particle's Cartesian coordinate vector in the forward 
run and ${\bf x}_b$ 
to the value of the corresponding coordinate in the reversed run,  
when the system is integrated 'backwards'. Errors $\xi = \xi (t)$  are thus obtained by subtracting  
forward-run coordinates at that time $t$ from the corresponding coordinates in the reversed run at $2 t_f - t$. 
For the simulations studied here $t_f = 100$ time units, which corresponds to about 29 
dynamical times at $r = a = 1$, 23 dynamical times at the half mass radius, and 16 dynamical times at the virial radius.  
  
Since when the dynamics is reversed the velocities are also reversed (and their Cartesian components change sign) the corresponding 
formula in this case is  
\begin{equation}
\xi_{\dot{x}}^2 = \frac {\sum_{N} |\dot{\bf{x}}_f +   \dot{\bf{x}}_b|^2} 
{\sum_N |\dot{\bf{x}}_b|^2}. 
\label{eq:relvel}
\end{equation}
Similarly for the angular momenta we have 
\begin{equation}
\xi_L^2 = \frac {\sum_N |{\bf L}_f +  {\bf L}_b|^2}{\sum_N |{\bf L}_b|^2}.
\label{eq:relang}
\end{equation}   
Finally, the errors in energies are calculated from.
\begin{equation}
\xi_E^2 = \frac {\sum_N (E_f - E_b)^2}{\sum_N E_b^2}.
\label{eq:relen}
\end{equation}

In what follows, we refer to these quantities as the `relative errors' in the various variables.   
We will also use the absolute RMS errors, these are obtained by replacing the denominators in 
the above equations by $N$.  In Section~\ref{sec:fixed}, when we examine the error growths in systems 
integrated with fixed stepsize, the errors will refer to differences 
between pairs of systems started with the same initial conditions and integrated forward in time 
with different (usually by a factor of two) stepsize.

\section{Propagation of errors}
\label{sec:main}

\subsection{Estimates of Systematic, diffusive and exponential error propagation}
\label{sec:sim_estimates}

In this section, we estimate the growth of errors in numerical integrations, relating it to the 
intrinsic dynamics of the stability of trajectories.

\subsubsection{Case I: No intrinsic divergence} 

If there is no intrinsic divergence between solutions of initial value problems started with differing 
initial conditions, then the only source leading to the separation of solutions in time 
will be numerical errors (e.g., this is the case for two numerical solutions
of harmonic oscillators  with the same spring constant).  If the errors at each step 
are added, they would generally lie between two extremes; they either add up to zero, so that
the RMS error is $\sim \sqrt{n}$ after $n$ steps, or are systematic in the sense that the errors 
scale as $n$ (e.g., Press et. al. 2007). 

\subsubsection{Case II:  Systems with phase mixing} 

A nonlinear dynamical system, even if it supports only regular trajectories with no exponential divergence, 
will generally exhibit phase mixing due to the dependence of the characteristic frequencies 
that describe it quasi-periodic motion on the amplitudes.  
If  such an integrable  system is Hamiltonian, its trajectories can always be parametrised 
in terms of action angle variables (${\bf J,  \Theta}$). When the dynamical evolution is exact (without numerical errors) 
the actions ${\bf J}$ are constant and the angles evolve as 
${\bf \Theta = \Theta}_0 + \mbox{\boldmath$\omega$} t$. Here,   
${\bf \Theta}_0$ comprises the initial phases at $t=0$, and $\mbox{\boldmath$\omega$}$ the frequencies
characterising the quasi-periodic motion. 

Since ${\bf J}$ does not evolve in time, 
the dynamics does not prompt any intrinsic divergence of
nearby solutions in these variables,  
and the numerical errors can again propagate 
systematically as $\delta J \sim n$ (where $\delta J$ is the magnitude of the 
typical error per timestep),  or diffuse as  $\delta J \sim \sqrt{n}$. 
The  angle variables, on the other hand, evolve linearly in time, so 
we expect the errors in them to propagate as
\begin{equation}
{\bf \delta \Theta} = \sum_{i = 1}^{n} \delta \mbox{\boldmath$\omega$}_i  \Delta t_i +  \sum_{i = 0}^{n} \delta {\bf \Theta}_i,
\label{eq:phase}
\end{equation}
 where $\delta \mbox{\boldmath$\omega$}_i$ and  $\delta {\bf \Theta}_i$ are, respectively, 
the numerical errors in frequencies and phases at timestep $i$.
Except for the zeroth term in the second sum, the error is expected to be dominated by 
truncation error rather than roundoff.  
As before, the truncation errors in that second sum can add up systematically to $\sim n~\delta \Theta$ or
diffusively as $\sqrt{n} \delta~\Theta$.
 However,  in addition to this,  the first sum displays a linear intrinsic divergence in time. 
If the numerical errors in $\omega$ add up systematically --- 
that is  $\delta \omega_i \sim i  \delta \omega$, for a 
typical error of magnitude $\delta  \omega$  associated with typical 
timestep $\Delta t$ ---  
the first sum becomes $\sim \delta \omega \Delta t \sum_{n}  i \sim n^2  \delta \omega \Delta t$. 
The resulting relative error 
is then $\frac{\delta \Theta}{\Theta} \sim n^2 \frac{\delta\omega \Delta t}{\omega t} = n \frac{\delta \omega}{\omega}$. 
Similarly, the error scales as $\frac{\delta \Theta}{\Theta} \sim \sqrt{n} \frac{\delta \omega}{\omega}$, if the single step errors are assumed to add up diffusively. 

Thus, despite the intrinsic linear divergence, nonlinear systems with regular trajectories, which are subject only to phase mixing, can be numerically integrated to arbitrary (machine dependent) precision, like their linear counterparts. 
For example, the errors in the Runge-Kutta method used in 
this study are fifth order in the timestep $\Delta t \sim t_f/n$.  They thus scale as 
$\sim (\Delta t)^5 \sim 1/n^5$. If the errors propagate at most linearly in $n$, 
then the total error after $n$ steps scales as $\sim 1/n^4$ or better, which clearly can be made arbitrarily 
small by decreasing the timestep. This will remain true even if we take into account that the phase space variables (coordinates and velocities) 
of regular trajectories do not necessarily diverge linearly in time as the angle variables do (since the former are expressed as Fourier series in the latter; e.g., BT). It will remain true 
as long as the divergence can be locally characterised everywhere as a low order power law 
(as we will see this will be the case once 
the exponential instability has saturated). 
Finally, for such systems, the errors in mean field conserved quantities, such as the energy and 
angular momenta, should propagate as those in the action variables. 
In particular, if they are primarily driven by independent 
two-body encounters, the 
propagation in time should be diffusive (that is, $\sim \sqrt{n}$). 
Due to the weak $n$-scaling, information loss arising from such diffusion is expected to be minimal.

\subsubsection{Case III: Exponential divergence} 
\label{sec:expomodIII}

In the presence of  exponential divergence the situation is quite different 
from  that outlined above.  Consider coordinates in the 6-$N$ dimensional phase space 
${ {\bf X} = ({\bf x}_1,..., {\bf x}_{3 N}, \dot{\bf x}_1, ..., \dot{\bf x}_{3 N})}$. 
A  small error 
$\delta {\bf X}_0 = \delta X_0~\hat{\bf e}_0$ 
in ${\bf X}$ at time $t_0$, becomes    
$\sim \delta X_0  e^{k \Delta t}~\hat{\bf e}_1$
at time $t_1= t_0 + \Delta t$.  Here,  
the unit vector $\hat{\bf e}_1$ is in the direction to which $\delta {\bf X}_0$ points after its  
transformation by the dynamics following a timestep $\Delta t$. This  
is not necessarily in the direction $\hat{\bf e}_0$, as  the dynamical phase space flow not only stretches but also rotates an error vector. This acts as to
 align it with a (dynamically changing) direction 
of maximum expansion, characterised by the exponentiation rate $k = 1/t_e$. Formally, 
when the infinite time limit is well defined, this is the maximal Lyapunov exponent 
(see e.g. Wolf et. al. 1985 for a lucid discussion).

 Each timestep is accompanied by a numerical (primarily truncation) error. 
In addition, the errors from previous timesteps are amplified by the dynamics, such that the final total error is a superposition of exponentially 
inflated errors. 
The total error after one timestep is thus 
$\delta X_0  e^{k \Delta t} \hat{\bf e}_1 + \delta {\bf X}_1$. Similarly, after a second timestep 
(assumed to be of the same size for simplicity), one has an error
$\sim \delta X_0  e^{2 k \Delta t} \hat{\bf e}_2 + \delta  X_1  e^{k \Delta t}  \hat{\bf e}^{'}_1 + \delta {\bf X}_2$. 
Again $\hat{\bf e}^{'}_1$ is not necessarily the same as $\hat{\bf e}_1$. However,  since the error unit vectors 
appearing in such expressions  are updated and superseded at each subsequent step,  after $n$ steps we can 
write 
\begin{equation}
\delta {\bf X}_n = \sum_{i=0}^{n} \delta X_i e^{k (n-i) \Delta t} \hat{\bf e}_i,
\label{eq:rangrov}
\end{equation}
by defining the unit vectors $\hat{\bf e}_i$ as those in
the direction to which the errors ${\bf \delta X}_i$, arising at step $i$, point after 
being transformed by the dynamics through $n-i$ timesteps.  
The exponential inflation implies that, for timescales large compared to the exponential time $t_e$, the errors propagate in a manner that makes it impossible to 
obtain arbitrarily accurate solutions by decreasing the timestep. 
 
Again one may suppose two extremes;  the $\hat{\bf e}_i$ can either be all in the 
same direction (no rotation), and so the elements of the sum add systematically, 
or they can be randomly directed. In the latter case they are 
likely to be normal for large $N$, as 
the RMS value of the cosine of the angle between two randomly pointing 
vectors ${\rm X}$ and ${\rm Y}$,  
$\cos ({\bf X}, {\bf Y}) = \frac{{\bf X} . {\bf Y}}{|{\bf X}| |{\bf Y}|}$, decreases as   $1/\sqrt{N}$ for large 
$N$.~\footnote{In terms of components in $d$-dimensions,   
$\cos ({\bf X}, {\bf Y}) = \frac{\sum_d  X_i Y_i}{\sqrt{\sum_d X_i^2} \sqrt{\sum_d Y_i^2}}$.
As this has the form of a correlator, if the components are uncorrelated random variables (as in the case of randomly oriented vectors),    
$\cos ({\bf X}, {\bf Y}) \rightarrow 0$ as $d$ increases, with `sampling noise' scaling as
$1/\sqrt{d}$. A formal proof of the increasing probability of orthogonality 
of high-$d$ vectors is given in Theorem 2.8 of  Blum, Hopcroff \& Kannan (2018).}
 
In this case one can write 
\begin{equation}
(\delta X)^2 = \sum_{i=0}^{n} (\delta X_i)^2 e^{2 k (n-i) \Delta t} \approx \frac{t_e}{2 \Delta t}  \Big(e^{2 t (n)/t_e} - 1\Big) (\delta X)^2, 
\label{eq:rangro}
\end{equation}
where the last approximate equality assumes a typical numerical error $\delta X$ per timestep, and a large $n$. 
As we will see below, the exponential instability is driven by increasingly 
close encounters,  as $N$ increases, with impact parameter that scales as 
$R/\sqrt{N}$ for a system of fixed total mass and characteristic size $R$. 
If one estimates the timescale of an encounter with impact parameter $b$
and characteristic speed $v$ as $\sim b/v$, then encounters crucial 
to the exponential growth of errors  take place on shorter timescales as $N$ increases
(as $v$ remains constant).   
The total number of encounters for all particles also increases with $N$. 
Thus, the direction of the error vector is expected to rapidly fluctuate for larger systems. 
Equation~(\ref{eq:rangro}) should therefore constitute an accurate estimate of the growth rate of  errors. 

In the numerical numerical studies presented below, the 
errors are calculated from the differences between the 
forward and reversed trajectories.  For time  $t  \ge t_f$, they are given by  
\begin{equation}
\xi^2 (t  -  t_f) = \Big |{\bf X} (t) - {\bf X} (2 t_f - t)\Big |^2 ~ \big  / ~ \Big |{\bf X} (t) \Big |^2.
\end{equation}
If these arise in the manner considered here, 
then, up to the saturation of the exponential instability,
they should be of the order of
\begin{equation}
\big(\delta X\big)^2 (n - n_f) = \sum_{i= n_f}^{n} \big ( \delta X_i \big )^2 e^{2 k (n - n_f  - i) \Delta t},  
\label{eq:rangrot}
\end{equation}
for $n \ge n_f$. Here $n_f$  is the typical number of steps in the forward run (corresponding to time $t_f$),
and the $\delta X_i$ represents  the relative error per time-step determined by the 
Runge-Kutta routine tolerance setting (as we discuss in the case of the velocities in 
Section~\ref{sec:comp_expo}).

\begin{figure*}

\centering
					                    
	\includegraphics[width=1.\columnwidth, trim = 0cm 0.cm 0cm 0cm, clip, angle = 0]{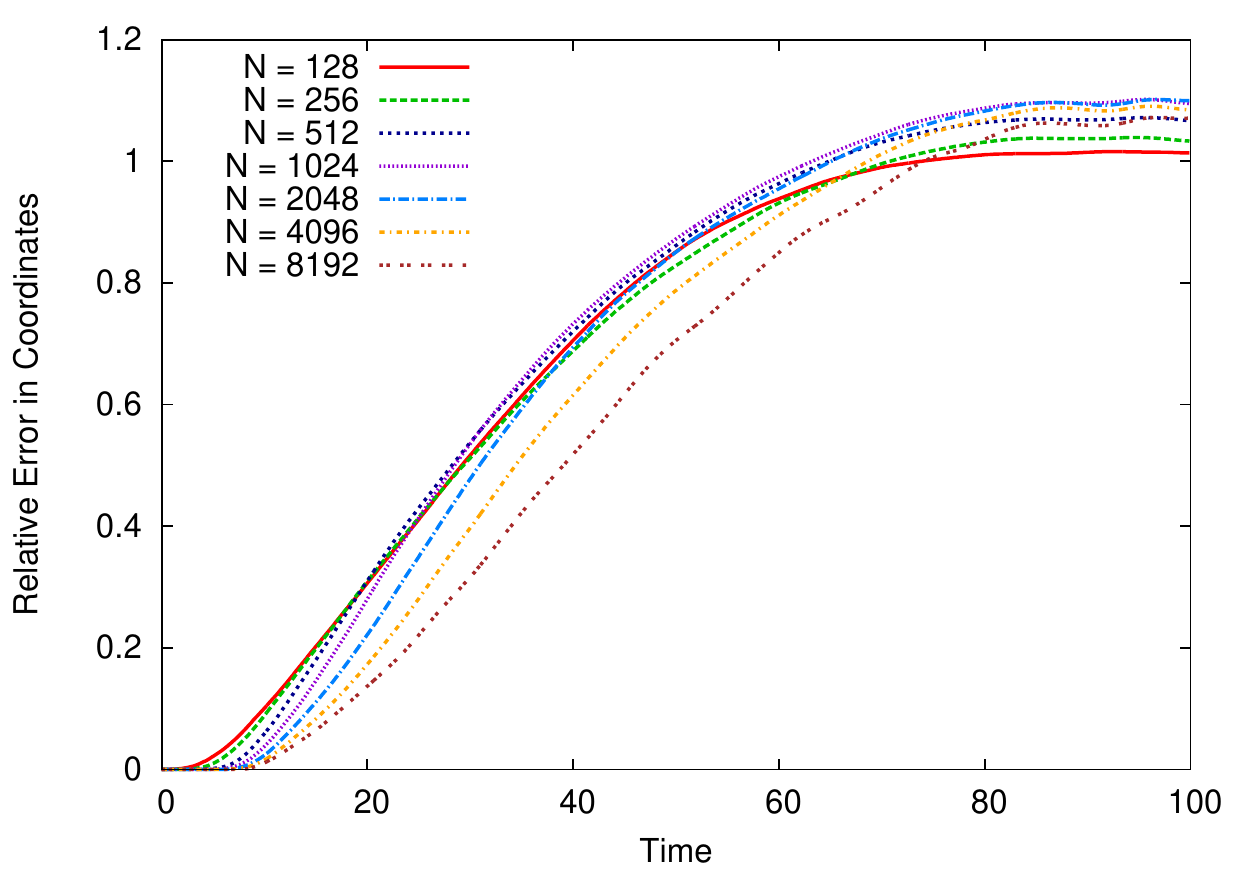}
	\includegraphics[width=1.\columnwidth, trim = 0cm 0.cm 0cm 0cm, clip, angle =0]{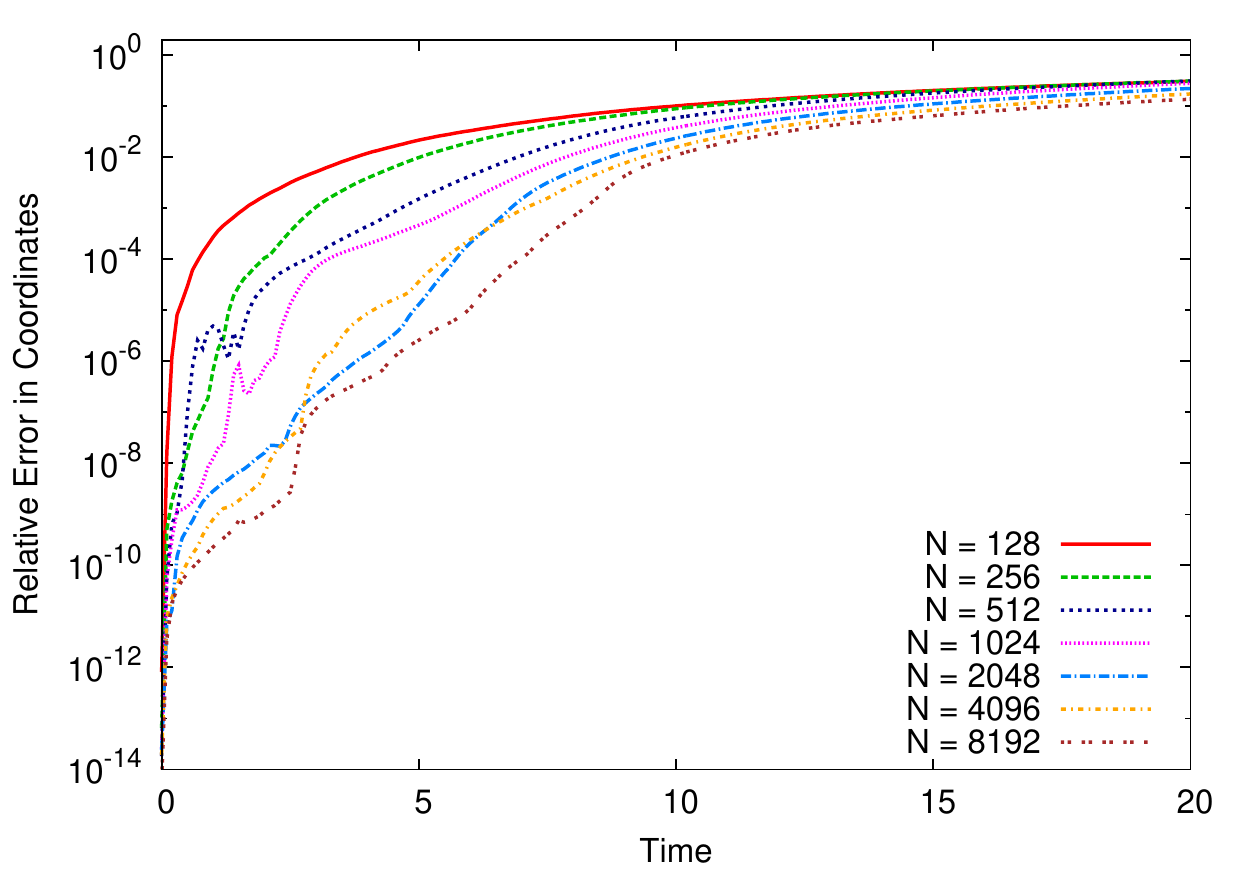}
						\caption{Growth of errors, measured from the difference in coordinates of particles in their forward  
and reversed trajectories, evaluated using equation (\ref{eq:relcoor}), 
as a function of time (measured from the start of the reversed run).  
The right panel shows the early growth with vertical 
log-scale. 
In the units used here, a dynamical time at $r = 1$ (i.e. at the core radius $a$ of equation~\ref{eq:Plum}) 
is $t_D (r= a) = 1/\sqrt{G \bar{\rho} (r= a)} = 3.4$; 
at the virial radius $t_D  (R_v) = 6.4$, and 4.4 at the half mass radius.}
    \label{fig:coorev}
\end{figure*}

\subsection{Propagation of errors in the phase space variables}
\label{sec:phase_space}

 In this section we discuss the propagation of errors in coordinates and velocities
in the numerical simulations we conducted. We interpret the results in terms of a model 
adapted from GHH.

\subsubsection{Coordinates}

The left panel of Fig.~\ref{fig:coorev} shows the relative error calculated from 
differences between the forward and reversed evolution in coordinates 
as defined in~(\ref{eq:relcoor}). The 
time is measured from the start of the reversed runs. On this linear scale, 
the errors are dominated by what appears 
to be quasi-linear growth, before flattening as the error becomes of order one, as correlation is lost between the 
coordinates of the forward and reversed systems.  This post-exponential growth is characterised by a  running power law index  
as we will see below (Fig.~\ref{fig:logderis}). 
 
The details of the early evolution are best examined with a vertical logarithmic scale  
(right panel  in  Fig.~\ref {fig:coorev}), which shows the growth of coordinate errors for the first 20 time units ($\sim 6~t_D (r=a)$). 
On this scale, linear sections correspond to exponential growth, while the growth 
during the first few time units corresponds to the quasi-linear addition of errors, implied by equation~(\ref{eq:rangrov}) on 
timescales smaller than the exponentiation time. 
The curves then flatten off, indicating a smaller exponentiation rate as the instability 
saturates. As $N$ increases, it saturates at smaller relative errors.

An important point to note is that this flattening takes place while the errors are orders of magnitudes smaller than one
(compare the values where the curves flatten on the linear and vertical scale plots).  
This need not be the case, as in many systems saturation of the exponential divergence of 
initially nearby trajectories occurs only when the 
separation is of order of the system size. A case in point pertains to chaos in smooth gravitational potentials, 
which can stem from global asymmetries  (such as triaxiality) in the smoothed out matter distribution. It
can be distinguished from `N-body chaos' 
by precisely this feature (e.g., Kandrup \& Sideris  2003). 
In our present case, relative errors at which saturation 
occurs here can be explained by a simple model due to GHH, which we now briefly describe
while adopting it to our purposes. More details are to be found in their original paper.  

\subsubsection{A simple model}
\label{sec:sim_mod}
\begin{figure*}

\centering
					                    
	\includegraphics[width=1.\columnwidth, trim = 0cm 0.cm 0cm 0cm, clip, angle = 0]{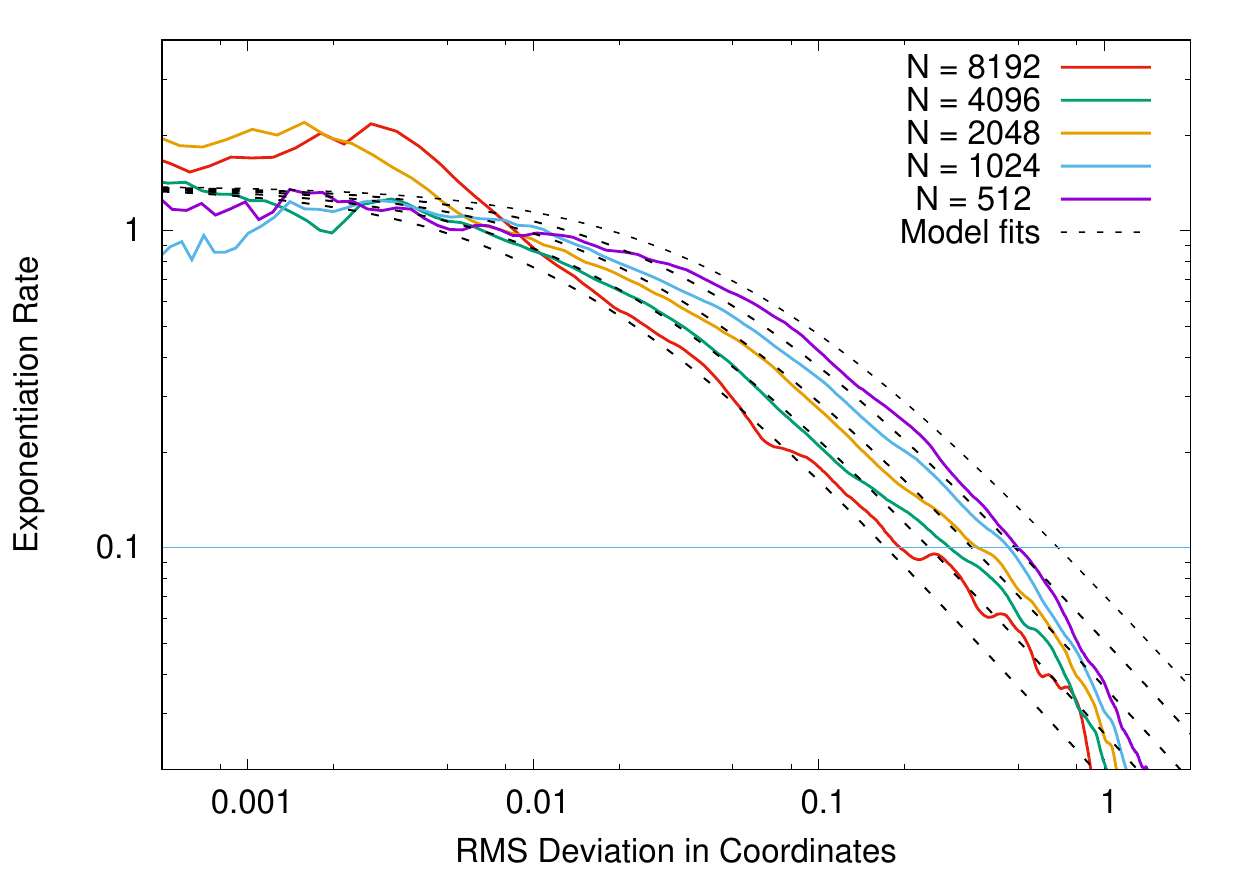}
	\includegraphics[width=1.\columnwidth, trim = 0cm 0.cm 0cm 0cm, clip, angle =0]{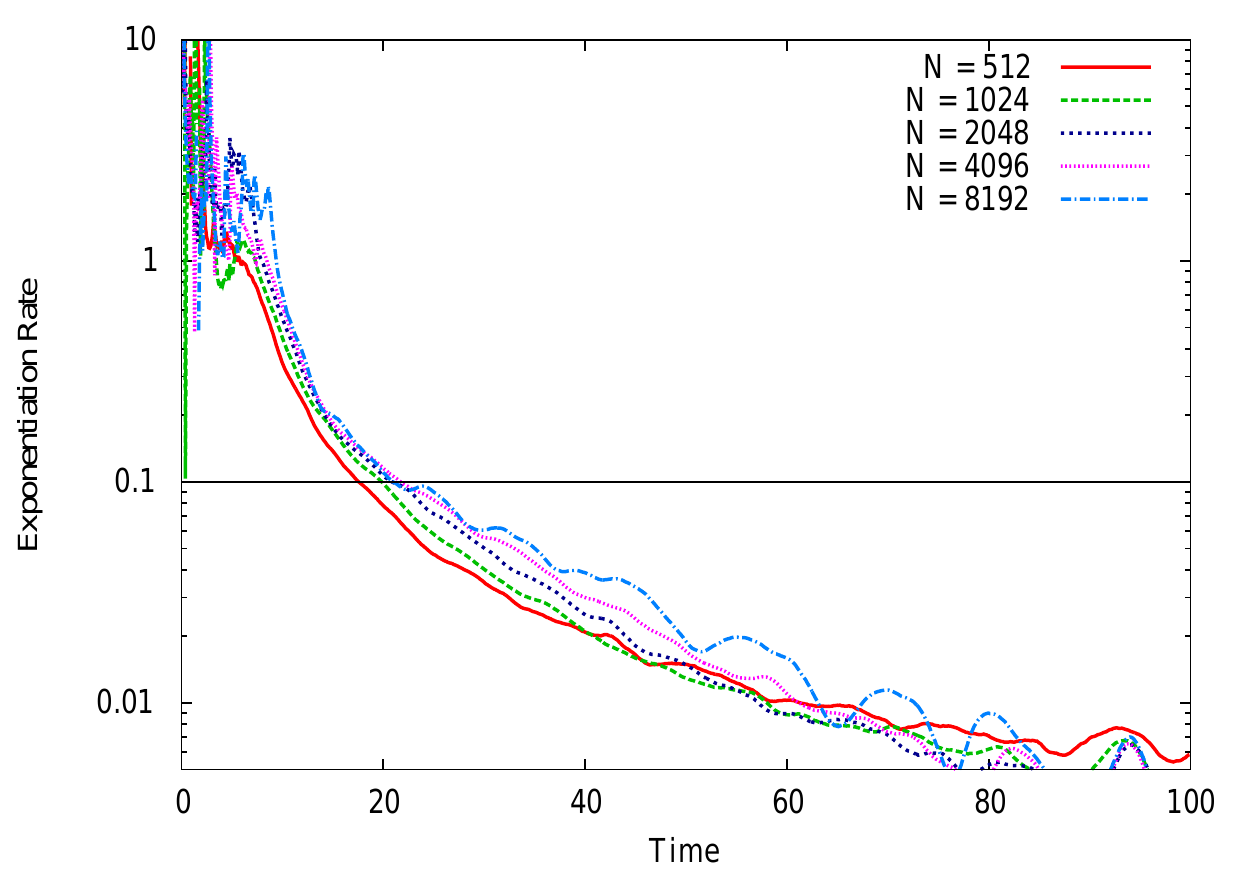}
\caption{Left panel: the numerical time derivative of the (natural) logarithm of the early growth of errors, 
which defines an `exponentiation rate', is compared with results obtained using equation~(\ref{eq:exratenl}) 
which describes coherent, multiplicative enhancement of error growth (dashed lines).  The 
 parameters used are 
$G M = \tau = v = 1$, $F R = 0.57$, in line with the characteristic mass and length scales of the simulated systems
(Section~\ref{sec:ICS}), and  our estimate  that the exponential divergence 
is driven by encounters with impact parameter $b \sim R/\sqrt{N}$, which occur roughly once per crossing time
per particle.  The parameters are fixed for all $N$, with $m = 1/N$. 
The theoretical results thus obtained are in 
 fairly good agreement with the numerical ones
(especially for relatively larger $N$), until the exponentiation rate drops below $k \la 0.1$.
From the right hand panel this can be seen to correspond to about $20$ time units. 
This corresponds to a period of systematic error growth ---   arising of the same multiplicative enhancement of errors that 
gives rise to the exponential phase --- that is different from simple phase  mixing. 
In fig.~\ref{fig:normcoors} we show the early growth of errors normalised to the expected saturation scale $0.57/\sqrt{N}$. 
It also illustrates the mechanism behind the worsening correspondence between 
the theoretical estimates and numerical results as $N$ decreases (as discussed in text).} 
    \label{fig:exrate}
\end{figure*}

Consider two test particles on nearby trajectories interacting with a third particle
of mass $m$. The impact parameters of the gravitational encounters involved are $b$ and  
$b + \delta b$.  
Assuming small deflection angles,  the difference in impact parameters
after time $\tau$ following the encounters is  $\delta b^{'} \sim  (1 +  G m \tau/ b^2 v) \delta b$, 
with $v$ the typical particle speed.  
If  the three particles involved
all move in a plane, it is clear that a second 
encounter at time $\tau$ will still increase the difference in impact parameters,  
such that $\delta b^{''} \sim  (1 +  G m \tau / b^2 v)^2 \delta b$, the sign of 
$\delta b^{'}$ notwithstanding.  If we assume that this is 
the case in general,  the effect of successive encounters can be multiplied  such 
that   $\delta b^{(l)} \sim  (1 +  G m \tau / b^2 v)^{(l)} \delta b$.
The associated exponentiation rate is then 
\begin{equation}
 k = \frac{1}{\tau} \ln \left(1 + \frac{G m \tau} {b^2 v} \right),
\label{eq:exrate}
\end{equation}
which is valid as long as $\delta b$ remains small. This simple picture is elaborated upon in
GHH, where it is shown to apply to encounters that drive the exponential instability (with 
typical impact parameters discussed below). 
Also, by combining the perturbations statistically, using the second moment equations, it is shown in their Section~3.2  
that the exponential growth persists
for (more distant) encounters for  
which the idea of simply multiplying the errors as above is not strictly valid.  
However, in this case the exponentiation rate is smaller 
and decreases with $N$.

\begin{figure}
\centering					                    
	\includegraphics[width=1.0 \columnwidth, trim = 0cm 0.cm 0cm 0cm, clip, angle = 0]{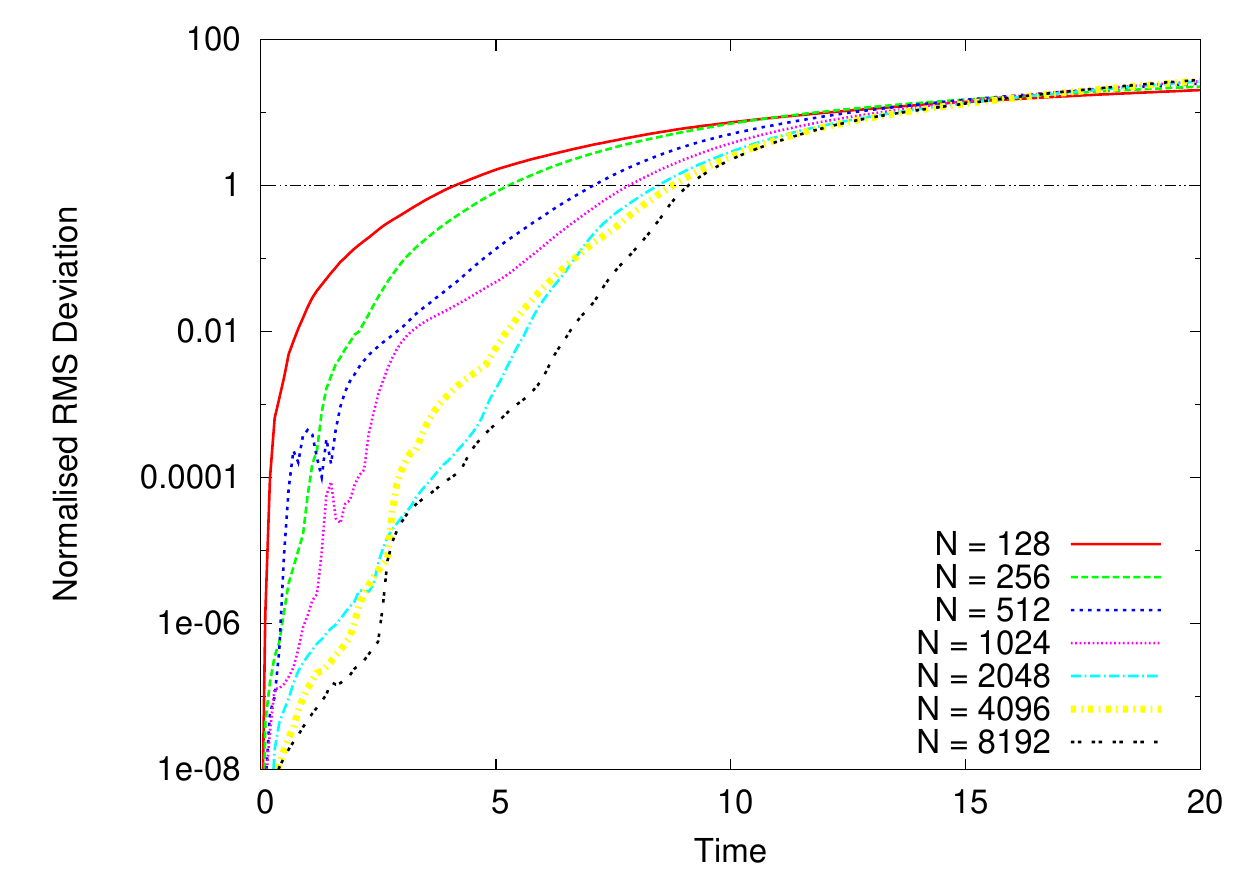}
\caption{Early growth of coordinate errors normalised to $0.57/\sqrt{N}$. For larger $N$ the curves 
flatten fairly sharply at a value of order $1$. For smaller particle numbers, the flattening is more gradual due to the 
rapid early growth, which implies that the errors are already large on timescales of order of the 
exponentiation time $t_e$ (i.e. at the start of the exponential divergence phase).} 
    \label{fig:normcoors}
\end{figure}

Now we note that if the separation  $\delta b$ is 
large, then during the encounter one has $\delta b^{'} =   b^{'}_1 -  b^{'}_2 =  (1 + G m / v b_1 b_2  ) (b_1 - b_2)$, so that for
$b_2 \gg b_1$,  $ b^{'}_2 = b_2$ (the particle much further away remains unaffected by the encounter). In this case,  we have 
$b^{'}_1 \sim b_1 + G m \tau / b_1 v$, and the  deflections add up to zero.  Though adding the squares of the velocity deflections, as appropriate for a  diffusive 
process, would lead to the standard two body relaxation time.  The exponential instability is thus intricately linked to the proximity of the two test 
particles relative to the impact parameter of encounter with the third particle.

\begin{figure*}

\centering
					                    
	\includegraphics[width=1.\columnwidth, trim = 0cm 0.cm 0cm 0cm, clip, angle = 0]{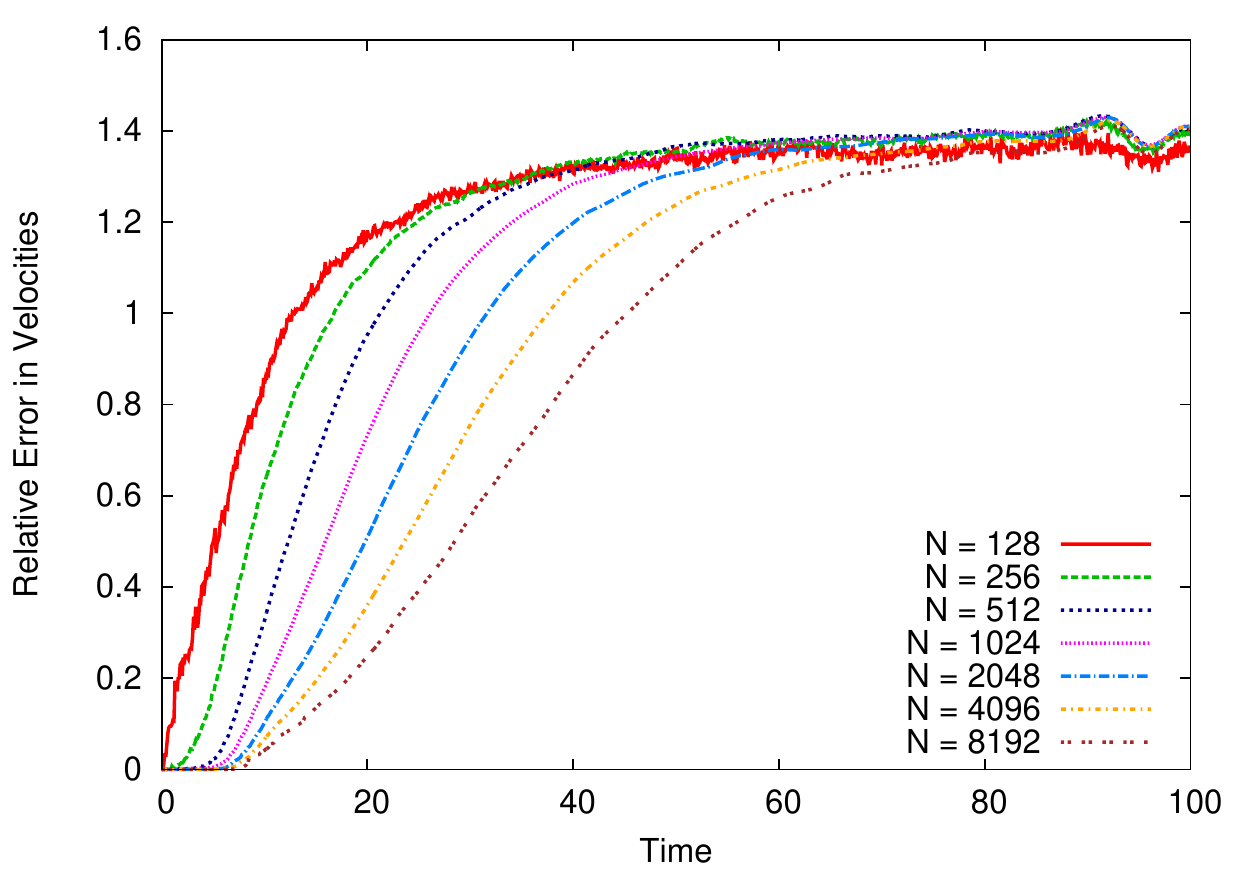}
	\includegraphics[width=1.\columnwidth, trim = 0cm 0.cm 0cm 0cm, clip, angle =0]{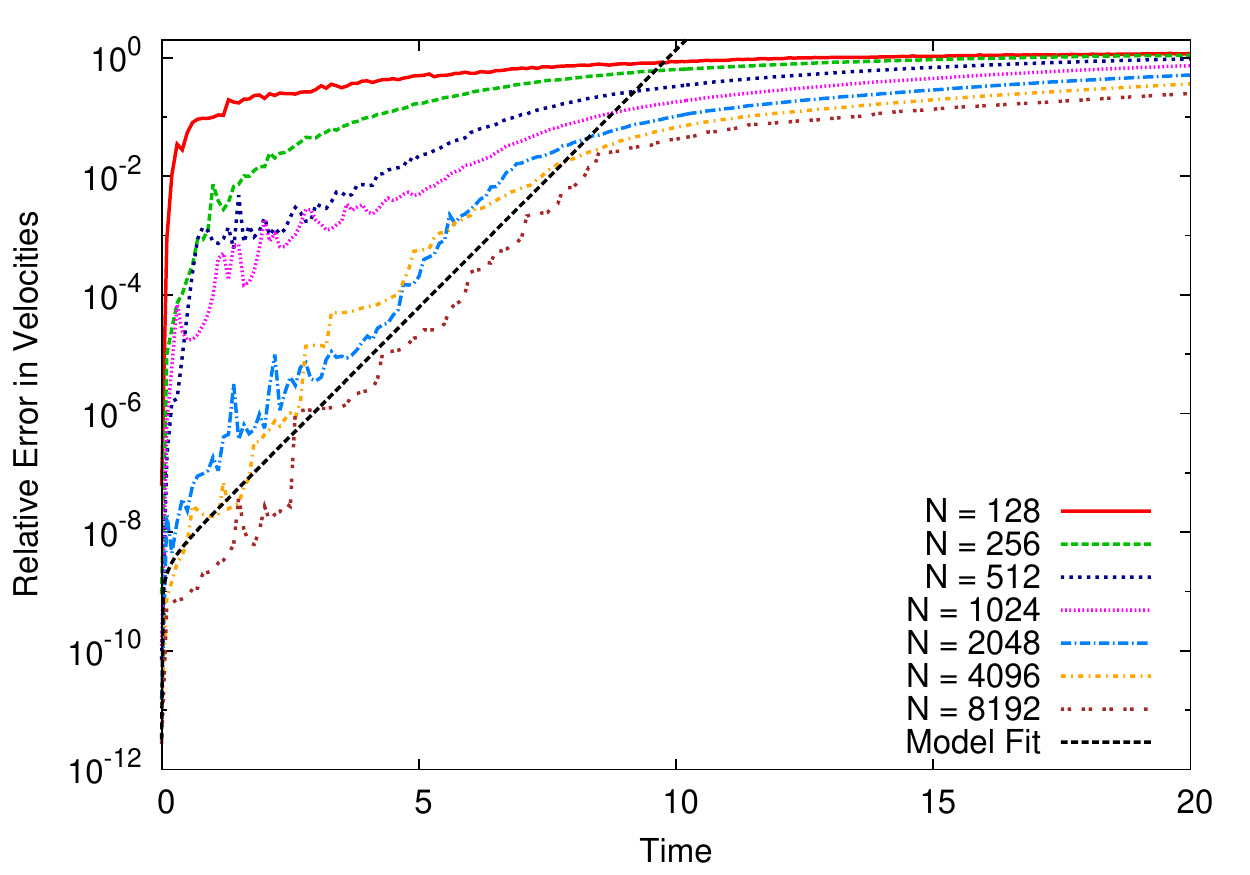}
						\caption{Same as in Fig.~\ref{fig:coorev} but for growth of velocity errors, 
calculated using (\ref{eq:relvel}).   The dashed line in the right hand panel 
 shows a theoretical estimate of the error growth using equation(\ref{eq:rangrot}), with assumed error per time-step of $10^{-9}$, hundred timesteps per dynamical time (at $r=a$), and exponentiation rate $k= 2$ (corresponding to an exponential timescale 
of $0.147 t_D (r = a)$).
} 
    \label{fig:velyev}
\end{figure*}

In this context, saturation of the exponential growth should 
occur when $\delta b$ can no longer be considered small
relative  to the characteristic impact parameter $b$ of encounters that drive the exponential 
growth. This, as opposed to saturation on the scale of the system size.  This 
is the major difference between the exponential instability in $N$-body systems and exponential instability
that implies 'global chaos'  and relaxation leading evolution in the system properties on the exponentiation scale.

 The $N$-scaling of the characteristic maximal impact parameter 
can be estimated by assuming that the exponentiation 
rate is  essentially $N$-independent.  This approximate $N$-independence   can 
be tentatively deduced from Fig.~\ref{fig:coorev} for larger $N$, 
and more precisely by solving the linearised variational equations
(as in, e.g., GHH
who propose a very weak $N$-scaling $\sim \log (\log N)$. Going to larger particle numbers, 
and solving directly the nonlinear equations of motion for initially nearby trajectories,  
Hemsendorf \& Merritt 2002 find a similarly weak dependence, with scaling 
$\sim \log (\log N)$ or  $\sim \ln A N$, with $A$ of order 1000.).
Under the assumption of  $N$-indepdence of the expoentiation rate 
we deduce the saturation spatial length-scale of the 
instability and test it against the numerical results.
 
  From equation (\ref{eq:exrate}), 
this exponentiation rate will be $N$-invariant --- when measured in dynamical times $t_D$, 
or in units that keep the total mass and characteristic size constant as used here ---  if both $\tau/t_D$ 
and  $\frac{G m \tau} {b^2 v}$ 
do not vary with $N$.  

To see when the first condition is satisfied we first note that,
for impact parameters $b$ and smaller, the number of field particles 
a test particle encounters while crossing a system of characteristic size $R$ are 
confined to a cylinder of volume $b^2  R$.  If their  number density is $N/R^3$, their 
total number within the cylinder is $b^2 R \times N/R^3$. 
Thus the number of encounters a particle 
undertakes while crossing the system is $\eta = t_D/\tau  \sim N b^2 / R^2$.
It is $N$-invariant (and so is $\eta^{-1} = \tau/t_D$),  if $b^2/R^2$ scales as $1/N$.
The second condition requires that  $\frac{G m \tau} {b^2 v}=  \frac{G M \tau} {N b^2 v}$ 
is $N$-independent.  Setting $G M \sim v^2 R$, $v \sim  R/t_D$ and   $\tau/t_D \sim R^2/N b^2$, as estimated above, 
one finds $\frac{G m \tau} {b^2 v} \sim \frac{R^4}{b^4 N^2} \sim \eta^2$, which is also constant 
if  $b^2/R^2$ scales as $1/N$.

Furthermore, if  $\eta \ll 1$,  encounters are too rare to be important. On 
 the other hand,  $\eta \gg 1$ corresponds to relatively large impact parameters, and
the statistical combination of such weak encounters lead to progressively smaller 
contributions to the exponential timescale (as shown in GHH).  Therefore, 
the main contribution to the exponential instability comes from encounters with $\eta \sim 1$. 
 That is, from encounters that occur 
about once per crossing time, and are characterised by impact parameters 
$b \sim \frac{R}{\sqrt{N}}$.

Thus, the exponential instability should start to saturate when the errors become larger than the typical 
impact parameter  
$\sim R/\sqrt{N}$. Which means that if we plot the exponentiation rate as a function of coordinate separation $\delta x$, it should be flat up to 
$\delta x \sim R/\sqrt{N}$, before falling off for larger $\delta x$. 
To describe this pattern quantitatively, we suppose that the errors remain multiplicative despite the relatively 
large separation.  In this case, we can replace $b^2$ in~(\ref{eq:exrate}) with $b_1$ and $b_2$, as the impact  parameters of 
diverging trajectories begin to differ significantly.  In that case 
 $k \rightarrow \frac{1}{\tau} \ln~[(1 + G m \tau / b_1 b_2) v]$. Furthermore, 
in accordance with our estimate  of the typical impact parameter driving the 
exponential instability, we can set $b_1 = F R/\sqrt{N}$ 
(where $F$ is of order one) and 
$b_2  \approx  b_1 + \delta x$.  To get 
\begin{equation}
k = \frac{1}{\tau} \ln  \left(1 + \frac{G m \tau / v}{F \frac{R}{\sqrt{N}} \left(F \frac{R}{\sqrt{N}} + \delta x\right)} \right). 
\label{eq:exratenl}
\end{equation}
In Fig.~\ref{fig:exrate} 
we compare the numerically derived exponentiation rate
with what is obtained using equation~(\ref{eq:exratenl}), with $\delta x$ derived from the numerically 
calculated RMS variation. As discussed in 
Section~\ref{sec:ICS}, the characteristic mass and length scales of our simulations 
are such that $G M = 1$ and $v \sim \sqrt{G M/R} \sim 1$,
if the characteristic system size $R$ is to be of the order of the Plummer sphere core $a = 1$. 
We thus set $G M = v = 1$. 
Furthermore, as the dynamical crossing time is also of order one, 
we set $\tau  = 1$, and finally $F R = 0.57$. The last two values are in   
line with the contention that the divergence between nearby 
trajectories is driven by encounters that occur roughly once  per particle per crossing
and with typical impact parameters $\sim R/\sqrt{N}$.   
These parameters are kept 
fixed for all $N$, with $m = 1/N$.  For exponentiation rate $k \ga 0.1$,
the results thus obtained are in good agreement with the corresponding numerical ones 
for relatively large $N$. Below, we briefly discuss the progressive worsening of the correspondence between the theoretical 
and numerical results as $N$ decreases, interpreting it in terms of the increasing role of strong encounters.

More generally, the gross features of Fig.~\ref{fig:exrate}
are straightforward to explain.   If  
the characteristic time for encounters driving the exponential divergence is 
$\tau \sim t_D \sim R/v$, 
and $v \sim G M/R$, the factor 
$G m \tau/ v$ in equation  (\ref{eq:exratenl}) is  $\sim R^2/N$ (given $m = M/N$). 
As $F$ is of order one, we have   
\begin{equation}
k \approx \frac{1}{\tau}  \ln  \left(1 + \frac{R N^{-1/2}}{(R N^{-1/2} + \delta x)}\right).
\end{equation}
Thus, as long as  $\delta x \ll R/N^{1/2}$, $k$ is approximately constant and $N$-independent. 
As $\delta x$ increases, however, $k \rightarrow 
\frac{1}{\tau} \ln (1 + R N^{-1/2} \delta x^{-1})$. And since, for large enough $N$ and non-infinitesimal $\delta x$,  
$R N^{-1/2} \delta x^{-1}$ is small,  then $k \approx  \frac{1}{\tau}  R N^{-1/2} \delta x^{-1}$. 
For exponentiation rate  $0.1 \la \ k \la 1$, 
this is also the approximate behaviour, including $N$-scaling of horizontal separation, of the numerically 
derived lines in Fig.~\ref{fig:exrate}.

 Thus, although the exponential timescale 
for the divergence of infinitesimally close trajectories does not depend on $N$, growth during the subsequent saturation 
stage does, and so does of course the total error after the exponential instability has saturated entirely.  
Also, the fact that the results obtained using~(\ref{eq:exratenl}) continue to follow the numerical ones 
even after the exponentiation rate $k$ has decreased by an order 
of magnitude, suggests that coherent scattering, 
leading to the multiplicative enhancement of errors in the manner 
assumed in (\ref{eq:exratenl}), remains effective even after the exponential growth stage proper has ended. 
This reflects a phase of systematic growth of errors in the phase space variables 
intermediate between the exponential (constant $k \sim 1$) stage and the regime where 
phase mixing finally dominates. 
One also expects this systematic error growth phase to separate an eventual diffusive 
evolution of errors in the mean field conserved quantities from the exponential growth stage. We will see that this is indeed the case.  

As $N$ decreases the results obtained using~(\ref{eq:exratenl})  
no longer follow the numerical ones as closely as they do for larger $N$. 
This is related to the fact that, for smaller $N$, the exponential  region 
of the growth of errors  is progressively smaller; early evolution is  dominated by the initial rapid growth, taking place on timescale smaller 
than the e-folding time, then saturation starts to set in. This  can be seen from the right hand panel of 
Fig.~\ref{fig:coorev} and fig.~\ref{fig:normcoors}, the latter showing the early growth in coordinate 
errors as normalised to the expected 
saturation scale.  It can also be clearly seen in the evolution of velocity errors, 
which we look at next.

\subsubsection{Velocities}
\label{sec:velyev}

The exponential growth in velocity errors (Fig.~\ref{fig:velyev}, right hand panel) saturates at the same time as the coordinates, 
but with larger values. This due to the steeper pre-exponential growth that sets the 
initial conditions for the exponential stage, and can also be explained in the context of the 
model described above; in its context (as in standard two-body relaxation theory), 
the velocity kicks are local, reflecting discontinuous jumps,  
while the errors in positions only subsequently grow as a result of these velocity changes (cf.  Equations~11 of GHH).  Eventually, the relative velocity errors reach a value of about $\sqrt{2}$, 
corresponding to complete loss of correlations between individual particle 
velocities in the forward and reversed  runs, while maintaining the same distribution.~\footnote{From 
equation~(\ref{eq:relvel}), the relative error 
$\xi_{\dot{x}}^2 = (\sum_{N} |\dot{\bf{x}}_f|^2  +  | \dot{\bf{x}}_b|^2 
+ 2 \dot{\bf{x}}_f  .  \dot{\bf{x}}_b) / 
\sum_N |\dot{\bf{x}}_b|^2$. The third term in the numerator becomes smaller as the forward and 
reversed motion velocity vectors decorrelate. If the velocities in the forward and reversed runs nevertheless have the same statistical distribution then 
$\sum_{N} |\dot{\bf{x}}_f|^2  \approx \sum_{N} |\dot{\bf{x}}_b|^2$. 
Thus $\xi_{\dot{x}}^2 \rightarrow 2$.}

For smaller $N$, the initial growth is dominated 
by strong encounters; from the right hand panel of Fig.~\ref{fig:velyev}, we observe
that there is hardly a well defined region of exponential growth in  
the case of $N = 128$, where there is very rapid growth to the 
$10 \%$ level, followed  
by saturation. 
This is due to the important contribution of large jumps in the velocities. 
In this case, one also expect  (as we will see below)
the accumulation of errors estimated by the adaptive Runge-Kutta routine to be smaller than those inferred from comparing forward and reversed trajectories. This is to be expected because  the former method considers, at each timestep, the same initial conditions (up to roundoff error), and estimates the local error incurred by undertaking this timestep. In the latter case, on the other hand, there is 
already a difference between the forward and backward initial conditions at the start of the encounter, 
due to differences accumulated through the previous evolution of each system. 
A strong encounter can strongly amplify this difference.

\subsubsection{Comparison with estimate of Section~\ref{sec:expomodIII}}
\label{sec:comp_expo}

The dashed line in the right hand panel of Fig.~\ref{fig:velyev} shows 
the initial rise of the growth of errors in  velocities --- for which, as we saw, 
the relative errors are large compared to those in the coordinates) ---
as predicted by the theoretical model of Section~\ref{sec:expomodIII}.  
For larger $N$, the error growth (before saturation sets in) follows the theoretical 
prediction, obtained using equation~(\ref{eq:rangrot}), 
with an assumed typical time-step of $10^{-2}~t_D (r = a)$ 
and a truncation error of  $10^{-9}$. 
This is smaller than the truncation error set by the relative tolerance used in the Runge-Kutta routine  ($Tol = 10^{-8}$). But 
 as this tolerance constrains the maximal error in any one of the phase space coordinates, while the estimate from the simulation is an RMS estimate, 
the two are consistent. This is not the case however with the low-$N$ runs. 
Here, the use of equation~(\ref{eq:rangrot}) requires 
assuming a tolerance level that is orders of magnitude 
above $10^{-8}$ if the theoretical prediction is to match the numerical results.  
Thus, for relatively small $N$, when contributions of strong encounters are non-negligible, the Runge-Kutta error estimate, coupled with our prescription for simple exponential growth,
underestimates the  error growth inferred from comparing the backward and forward trajectories. 

\begin{figure*}

\centering
					                    
 \includegraphics[width=1.\columnwidth, trim = 0cm 0.cm 0cm 0cm, clip, angle = 0]{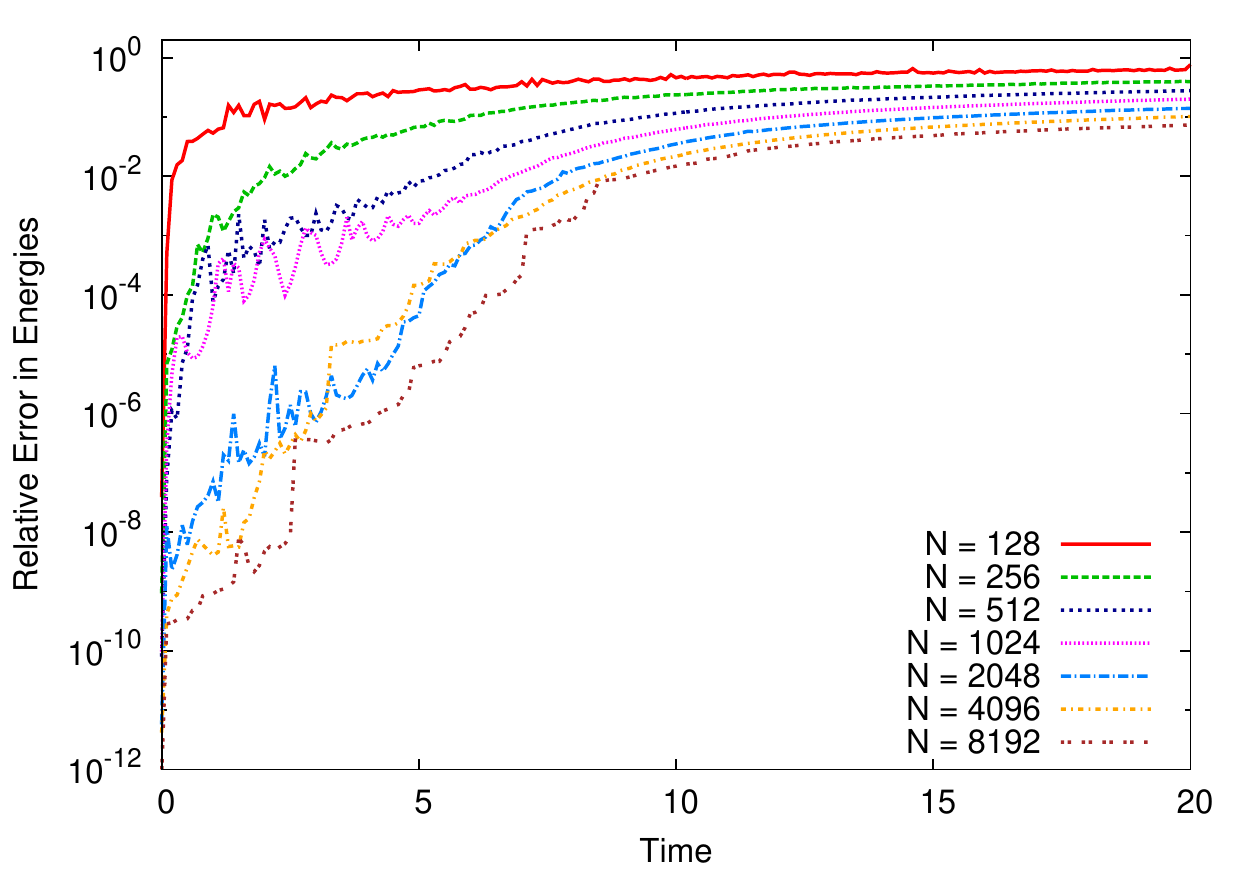}
\includegraphics[width=1.\columnwidth, trim = 0cm 0.cm 0cm 0cm, clip, angle =0]{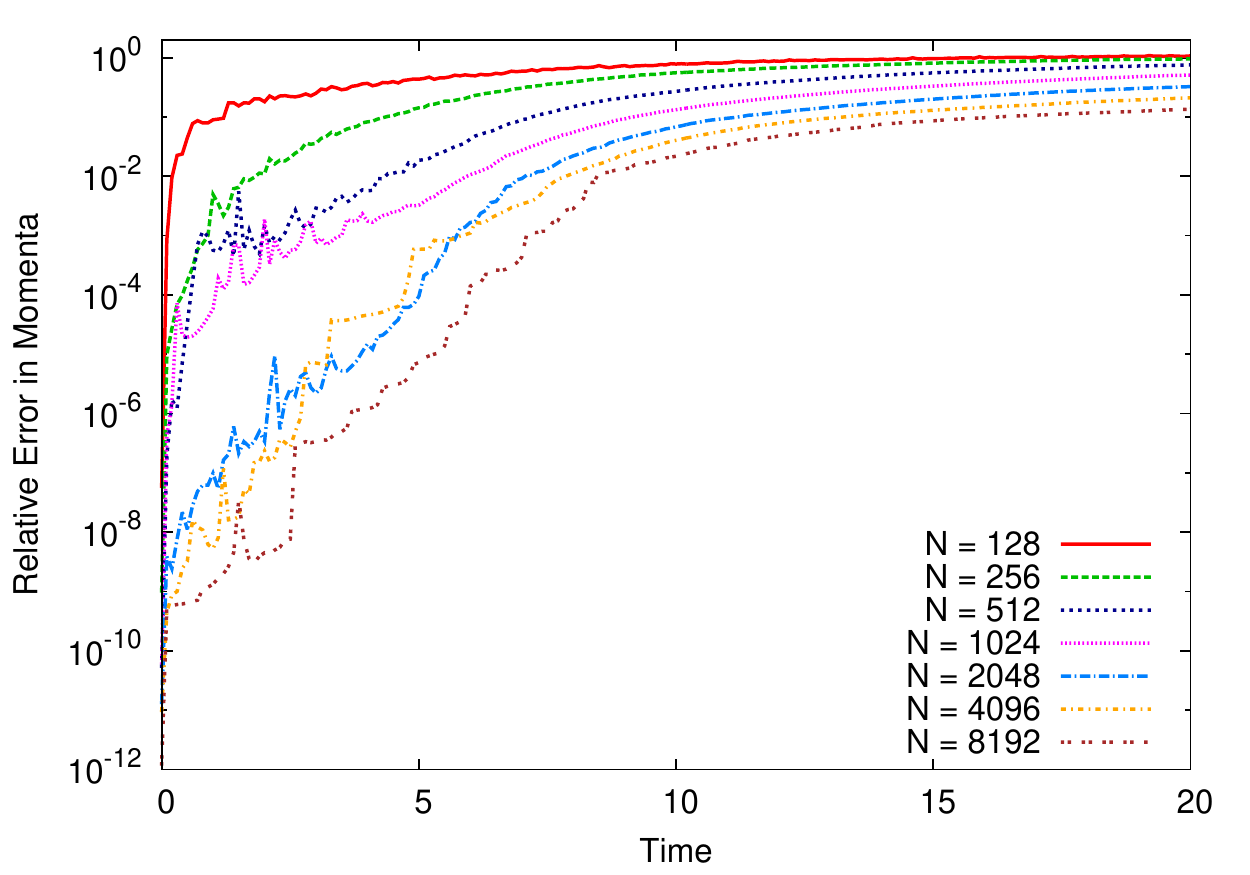}
						\caption{Early growth of the relative error in the mean field conserved quantities;
the energy (left) and angular momenta (right), calculated using equations  (\ref{eq:relen}) 
and~(\ref{eq:relang}) respectively. This early error 
growth is  similar in magnitude to that of the velocities, with about the same exponential 
growth rate as both coordinates and velocities; the exponential instability thus does not distinguish between the phase space variables and quantities that are conserved in the mean field, 
collisionless, limit. 
The flattening that follows the exponential stage is sharper in the case of energy however 
(and angular momentum for larger $N$).}

\label{fig:early_int}
\end{figure*}

\begin{figure*}

\centering
					                    
\includegraphics[width=1.\columnwidth, trim = 0cm 0.cm 0cm 0cm, clip, angle = 0]{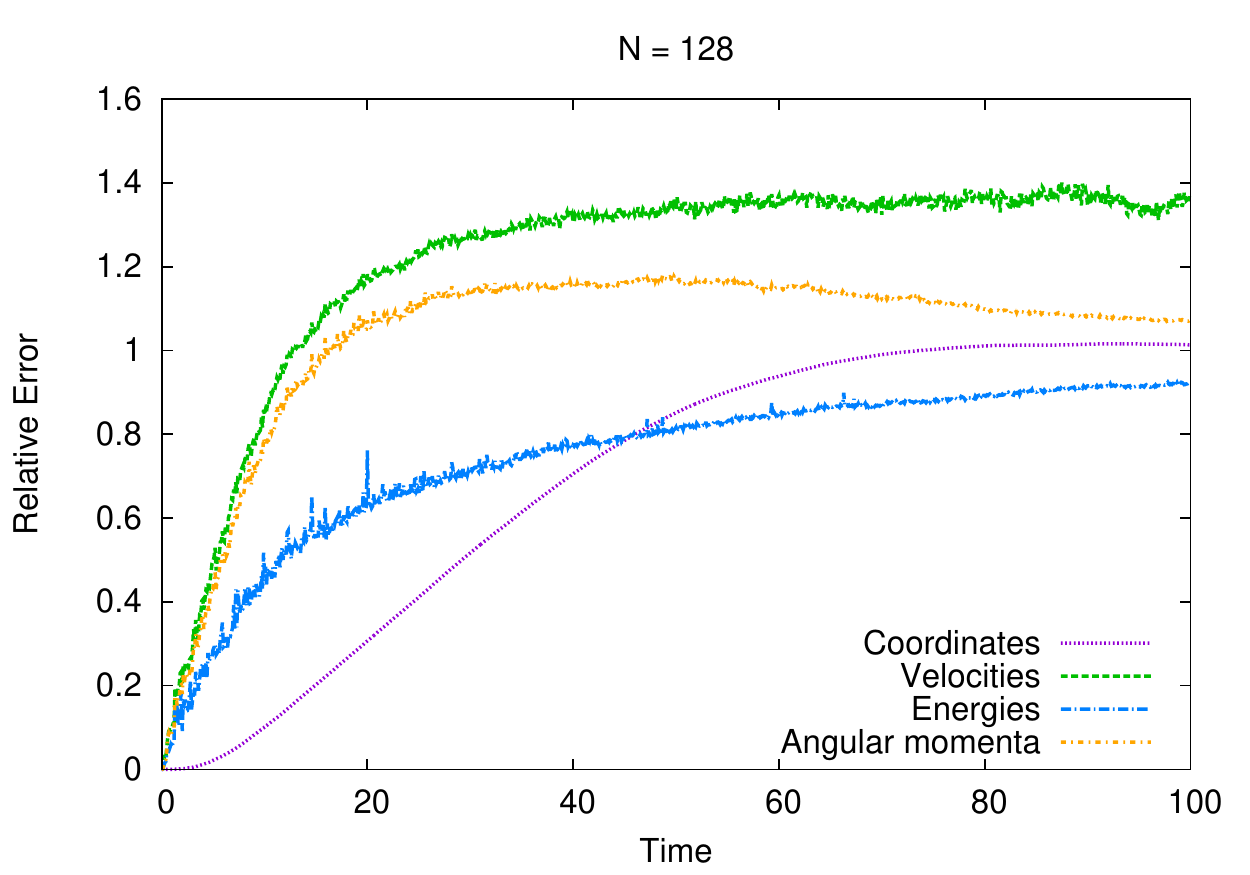}
\includegraphics[width=1.\columnwidth, trim = 0cm 0.cm 0cm 0cm, clip, angle =0]{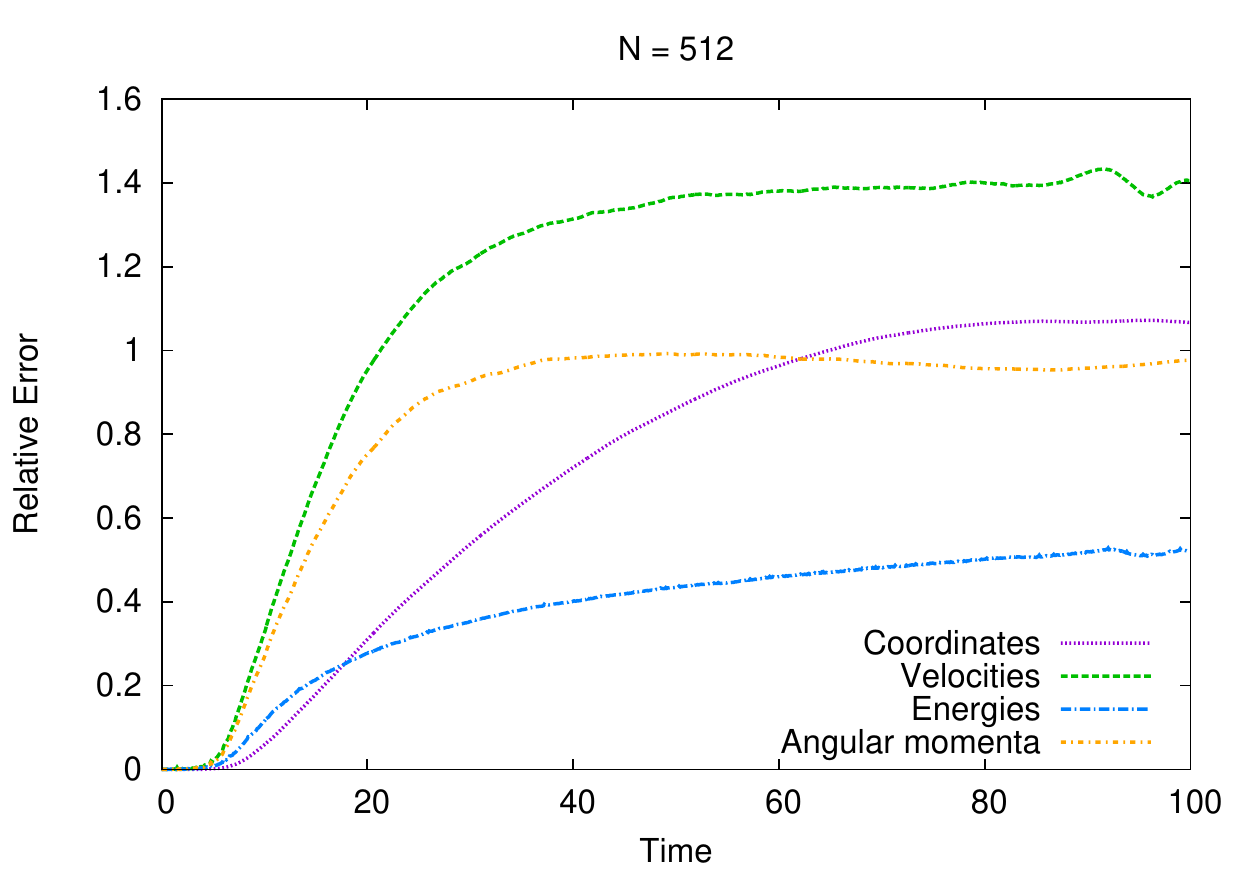}
\includegraphics[width=1.\columnwidth, trim = 0cm 0.cm 0cm 0cm, clip, angle = 0]{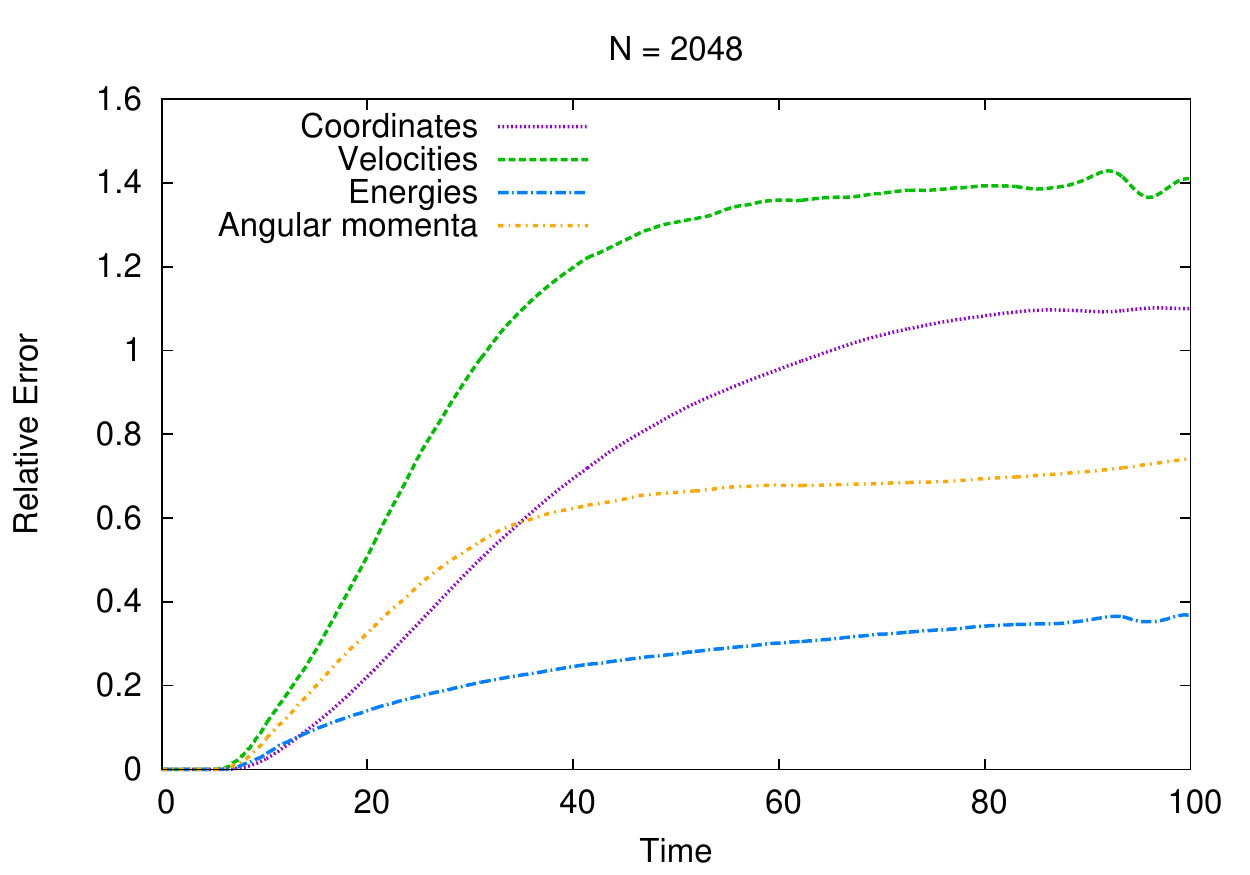}
\includegraphics[width=1.\columnwidth, trim = 0cm 0.cm 0cm 0cm, clip, angle =0]{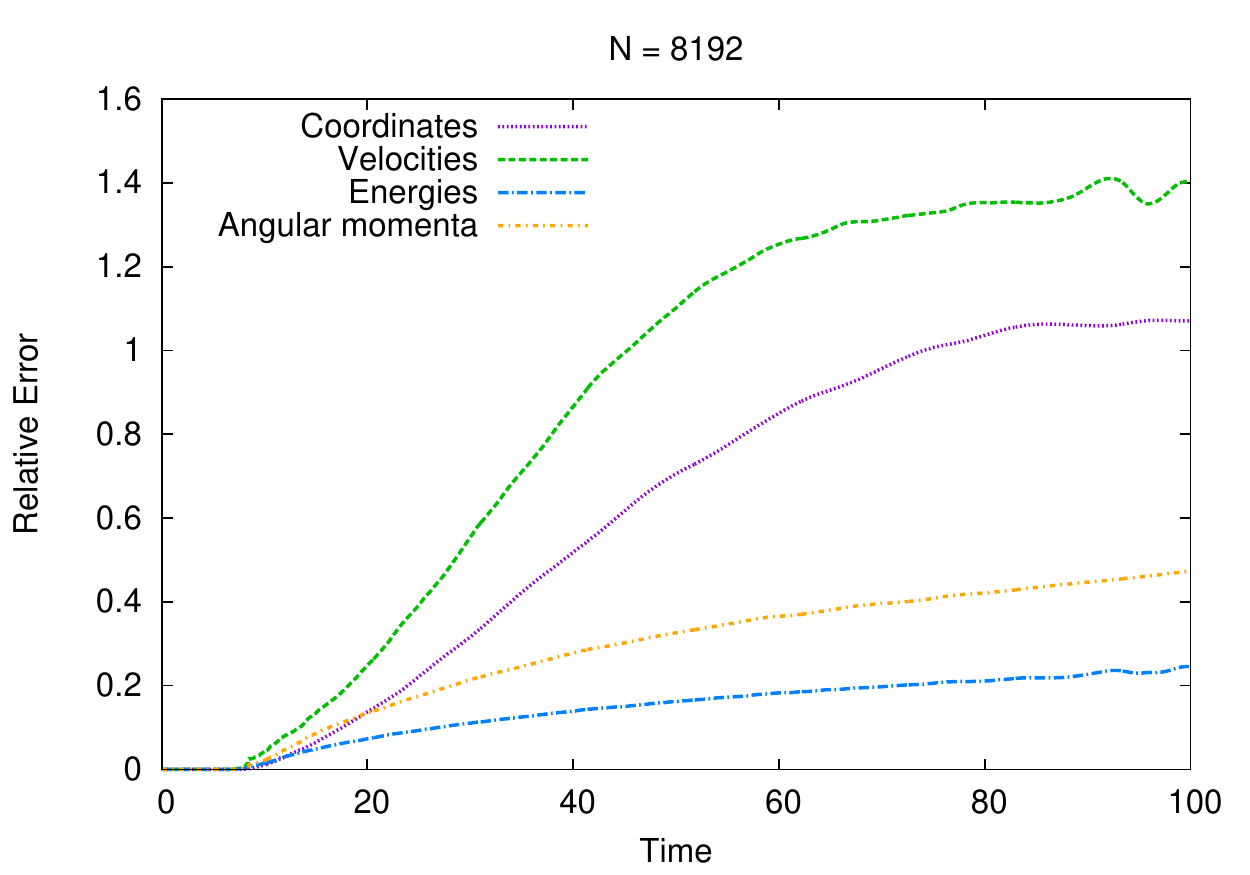}
\caption{Comparison of the error growth for $N= 128$, $N = 512$, $N = 2048$ and $N = 8192$. Note 
that  the growth in angular momentum mimics that of the velocities early on, before the former saturates.  
Beyond this point, the  velocity errors grow through phase mixing, while 
momentum errors grow diffusively. This diffusive growth is only clearly seen 
for relatively large $N$ and large times; for lower $N$, the  errors from the previous stages are large 
and the diffusive error growth is drowned out by this background.  
For energy, beyond $t \ga 30$, diffusive growth of errors is always apparent
(see also Fig.~\ref{fig:logderis} and associated discussion of the approach to the diffusion limit).}

\label{fig:allvar}
\end{figure*}

\begin{figure}

\centering
					                    
	\includegraphics[width=1.\columnwidth, trim = 0cm 0.cm 0cm 0cm, clip, angle = 0]{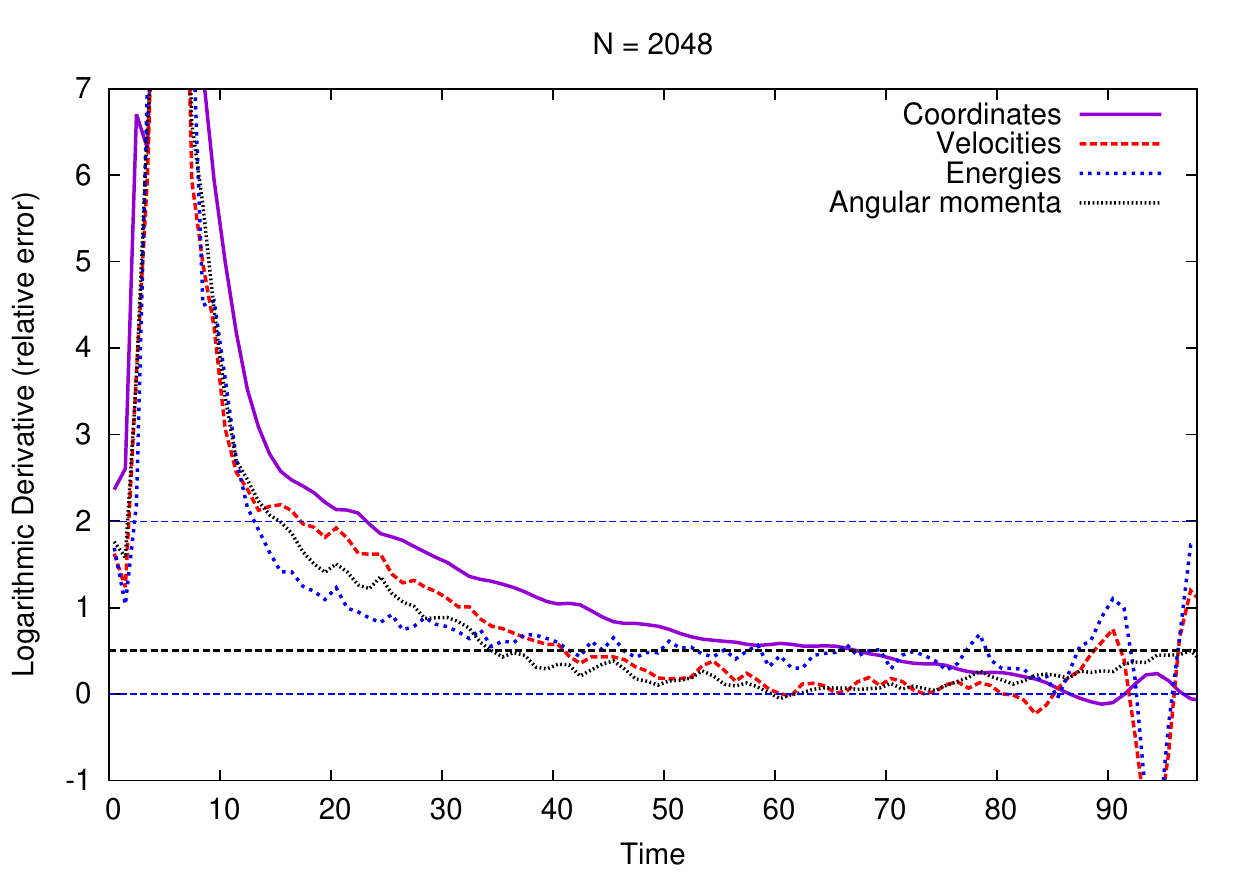}
	\includegraphics[width=1.\columnwidth, trim = 0cm 0.cm 0cm 0cm, clip, angle = 0]{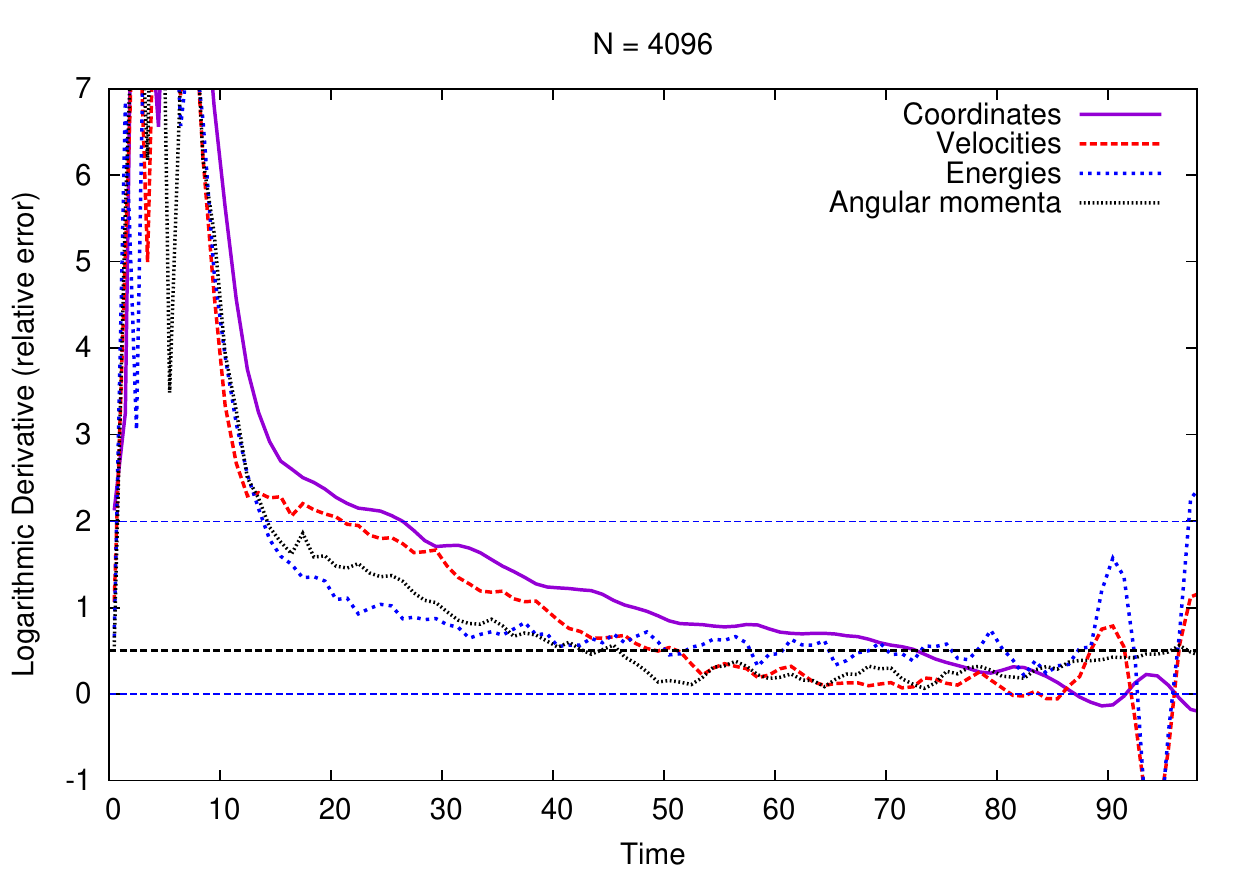}
        \includegraphics[width=1.\columnwidth, trim = 0cm 0.cm 0cm 0cm, clip, angle = 0]{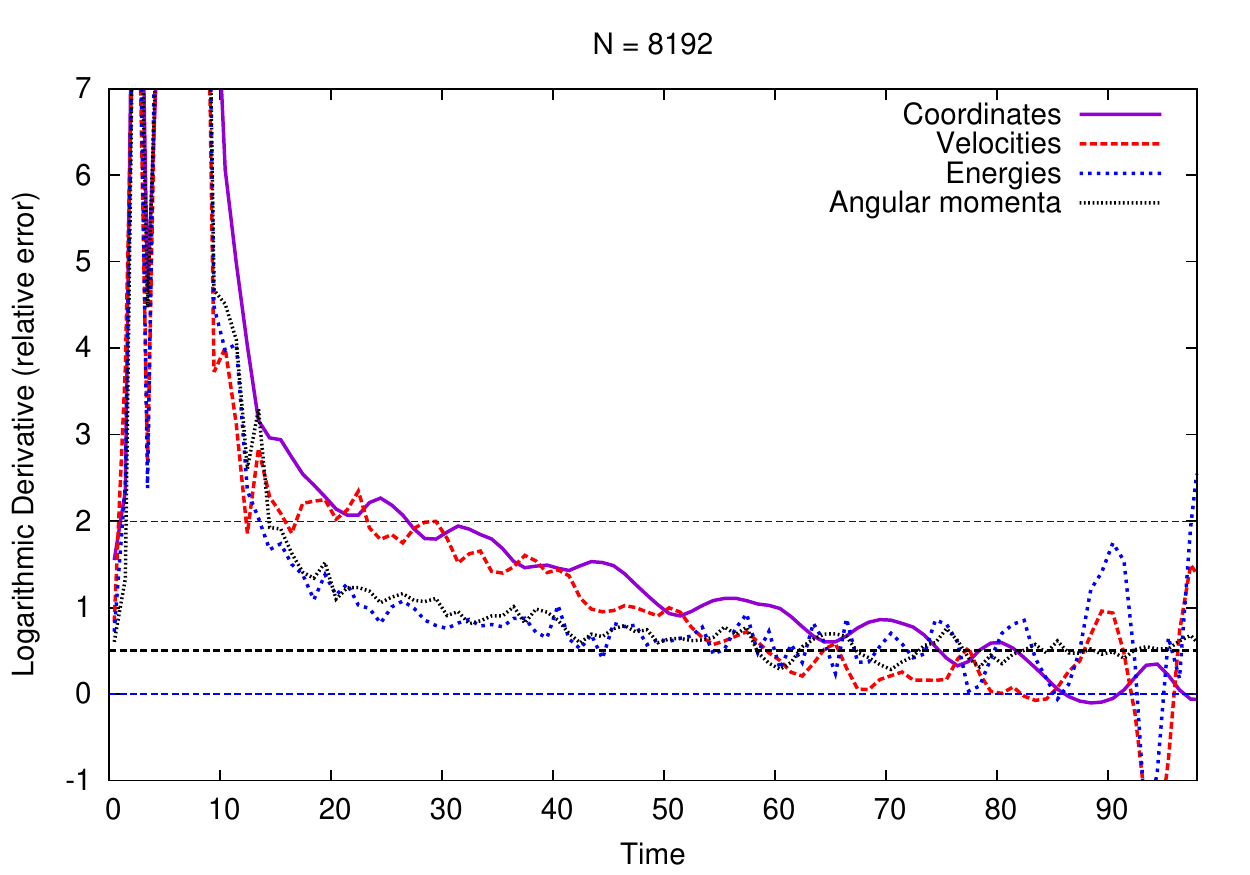}
						\caption{Logarithmic derivatives of error growths, for $N= 2048, 4096$ and
$8192$. The early rise corresponds  to stages of additive and  exponential 
inflation of errors; the rapid fall and initial flattening (to $t \sim 20$) roughly corresponds to the period through which 
systematic error growth can still be modelled mainly in terms of multiplicative enhancement by successive encounters. 
Beyond this, the errors in the phase space variables and the mean field conserved quantities grow qualitatively differently.
The former may still grow systematically {\it via} phase mixing, while the latter tend toward the diffusive limit, that is 
$\sim t^{1/2}$ and logarithmic derivative of $0.5$.   For relatively small $N$, the diffusive 
error growth in angular momentum 
becomes apparent only at the end of the timescales shown.}
    \label{fig:logderis}
\end{figure}

\begin{figure*}

\centering
					                    
	\includegraphics[width=1.\columnwidth, trim = 0cm 0.cm 0cm 0cm, clip, angle = 0]{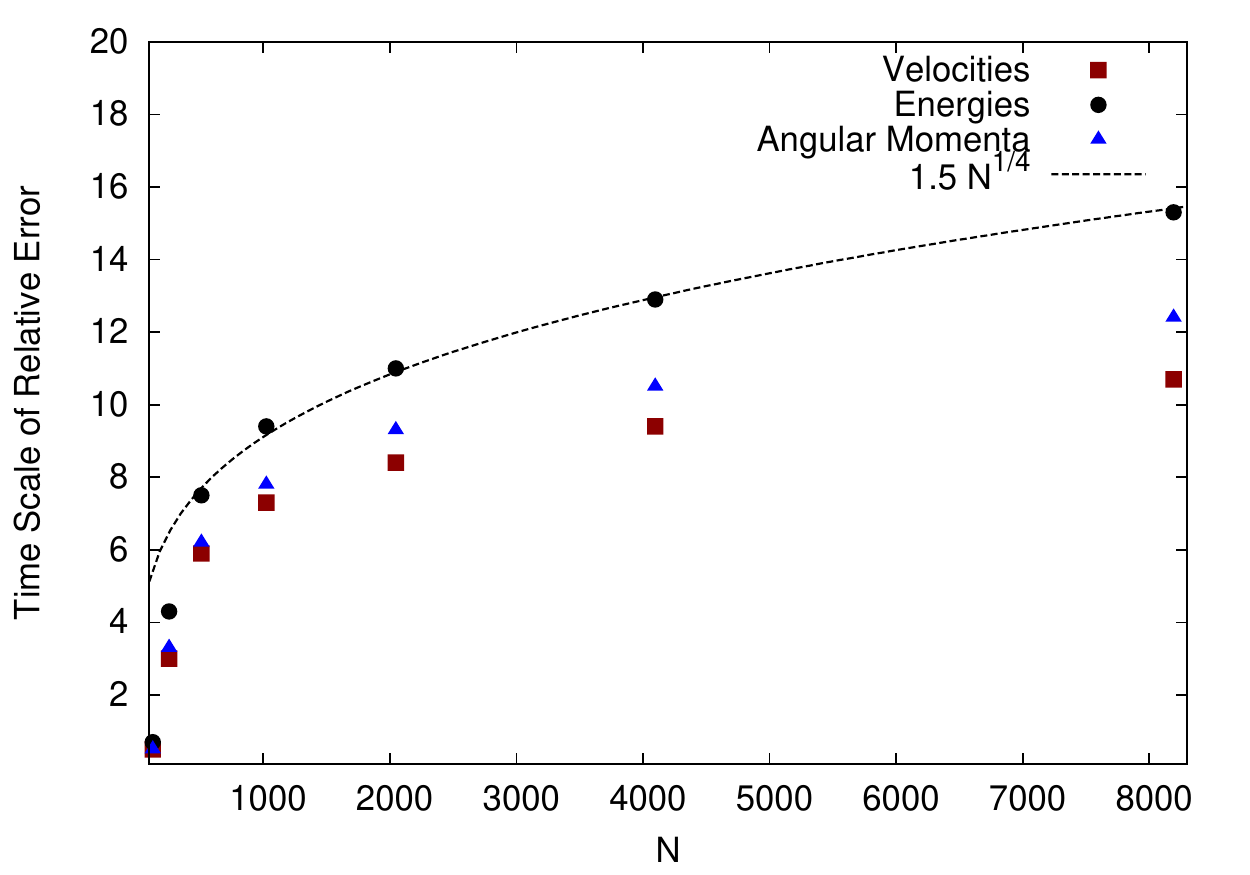}
	\includegraphics[width=1.\columnwidth, trim = 0cm 0.cm 0cm 0cm, clip, angle = 0]{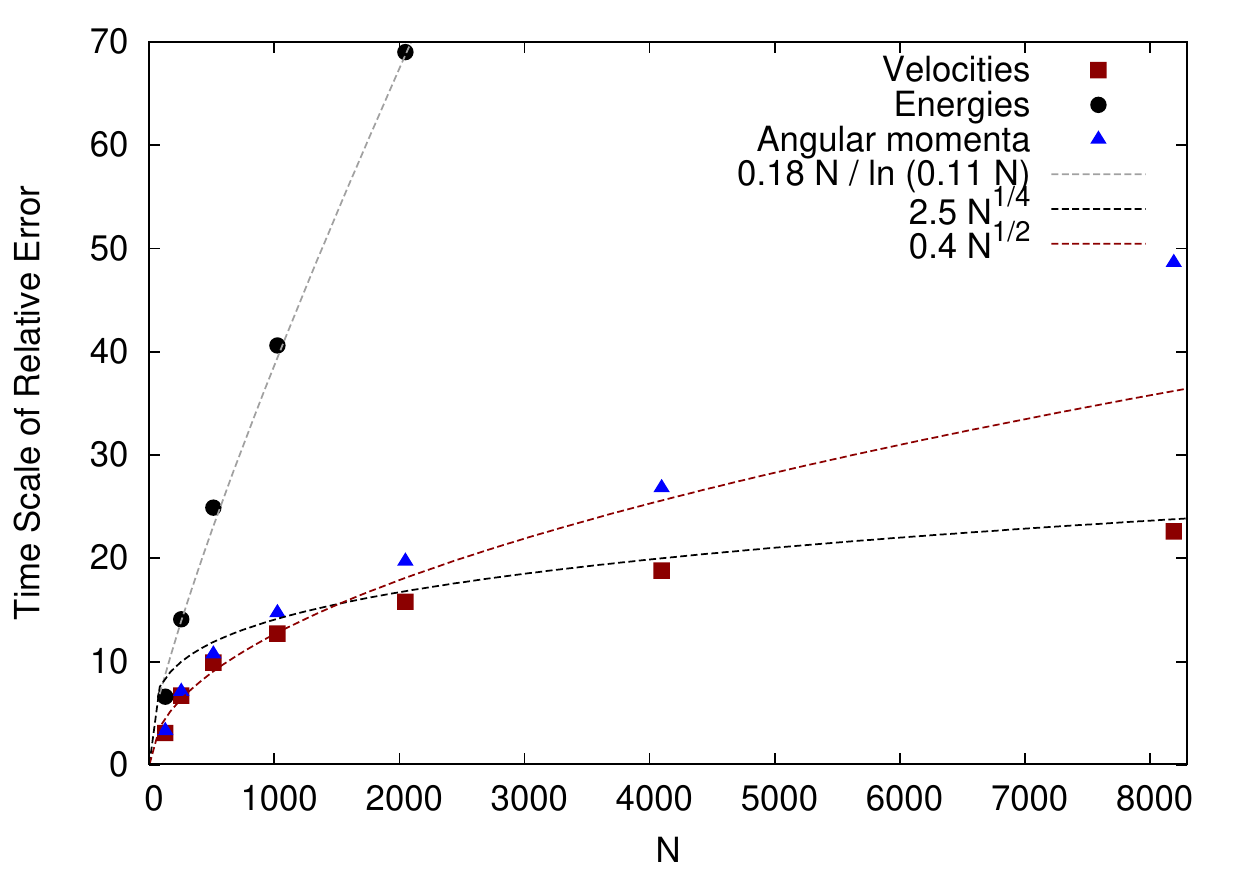}
						\caption{`Relaxation times' determined by evaluating differences between the forward and reversed 
trajectories at the $5 \%$ (left) and $33 \%$ threshold levels. In the former case, the inferred timescales are similar for velocity 
and angular momentum errors, and also for energy at lower $N$. At the $33 \%$ error level, the energy relaxation time follows the familiar two-body relaxation form. To  compare with the standard 
estimate, we can extrapolate (assuming diffusive evolution) up to $100 \%$ error level. 
In terms of dynamical time, the eye-fit to the energy error growth shown then gives
$t_{\rm relax}  = 0.26 N/ \ln (0.11 N) t_D (R_v)$.  For smaller  $N$, the velocity and 
angular momentum  errors grow in a similar manner up to the $33 \%$ threshold (as expected from Fig.~\ref{fig:allvar}), before the  effect of the diffusive character of the angular momentum error growth becomes apparent. 
The systematic (ballistic, as opposed to diffusive) growth in time, characteristic of this evolution, can be parametrised 
by power law index $1 \la s  \la 2$.}
\label{fig:Thresh}
\end{figure*}

\subsection{Errors  in mean field conserved quantities}

The growth of errors described above can be characterised by three stages; an early quasi-linear addition of errors, a  period of exponential growth, and an extended period of slower increase. 
The first two stages of error growths in energy and angular momentum, 
which are conserved integrals of motion in 
mean field equilibrium systems, are similar to the phase space variables (as can 
be seen from Fig.~\ref{fig:early_int}).  

In the third and final phase of error growth, the errors in the phase space 
variables are expected to grow  through phase mixing., a process which should
not affect conserved quantities. 
The error in these quantities may be expected to grow
diffusively if its growth follows the standard picture of two body relaxation.   
Although this is eventually the case, the angular momentum error growth 
in particular shows that this is not the whole story, as there is an intermediate 
stage of systematic error growth that neither corresponds to phase mixing or 
diffusion.  

The angular momentum error growth 
follows that of the velocities  well into the post-exponential regime. 
This is especially true 
for relatively small $N$, as can be seen from Fig.~\ref{fig:allvar}. 
It suggests that the effect of coherent, multiplicative, scattering
persists well beyond the exponential stage of constant exponentiation time, as can already be 
inferred from Fig~\ref{fig:exrate}. 
This causes systematic, as opposed to diffusive, growth in errors.   
For angular momenta, a proper diffusion limit is, in fact, only 
clearly apparent for the case with $N = 8192$, when the systematic growth saturates at small enough values as to allow for the subsequent 
diffusive evolution in errors to  become apparent over the timescale considered here.  
 In Section~ \ref{sec:fixed} we will see that the diffusive stage can be clearly detected also for smaller $N$ (but, 
still, for larger thresholds than the energy), when using a 
fixed timestep, as the use of an adaptive timestep seems to  enhance 
the effect of the instability spurred by coherent multiplicative scattering on the systematic 
growth of angular momentum errors at smaller $N$. In all cases however, a systematic phase of error growth separates 
the exponential from the diffusive evolution, such that the latter is only reached for relatively large error 
thresholds as we will see further on (sections~\ref{sec:relax} and \ref{sec:fixed}; particularly figures \ref{fig:Thresh}
and \ref{fig:Thresh-2}).

The panels of Fig.~\ref{fig:allvar} show that the error growth rate in energy, which is the usual benchmark employed to test for 
the presence of relaxation and validity of the collisionless limit, is slower than in the angular momenta. 
The precise mechanism behind this phenomenon may well be worth studying in detail, but is beyond our present scope
 (though, again, the results of Section~\ref{sec:fixed} suggest that part, but not all,  of the discrepancy is due to the use of a variable timestep). 
We mention, nevertheless, that error growth in energy 
may be heuristically understood in terms of  it being a scalar. 
Consider, for instance, the following  simple example: 
two particles  are
deflected by a third one (sharing the same plane), such that the respective deflections lead to normal velocity perturbations 
$v_1$ and $v_2$. 
Next, consider another encounter with the same 
magnitude of impact parameters and relative velocity, but where  the perturbing particle is on the opposite side 
of the two test particles; 
if  particle 1 and 2 were at impact parameters $b_1 < b_2$ from perturbing particle during the first encounter, 
they are at distances 
$- b_2$ and $- b_1$ respectively, in the second one. The 
perturbations to the velocities of the first particle is now $v_1 - v_2$, and that of the second particle is $v_2 - v_1$. 
The vector difference is $2 (v_1 - v_2)$. This will affect the difference in angular momenta. The kinetic energy
of each particle also changes due to the two 
perturbations, but the net difference after the two encounters is zero. In terms of the GHH model, it would appear 
that it is mainly the weaker encounters, which add up statistically, that would contribute in the case of energy divergence, 
leading to a clearer transition to the diffusion limit at smaller error thresholds.

To further quantify our statements concerning the stages of evolution of error growth, we have calculated the logarithmic 
derivatives of the errors. These are are presented in Fig.~\ref{fig:logderis}, where a vertical value of $0.5$   corresponds to 
$\sim t^{1/2}$ diffusive growth. As can be seen, this is clearly reached for the energy, but even then only after significant  
post-exponential 
evolution. (In  these plots, the early rise corresponds to exponential stage, while
the subsequent rapid fall and flattening  
up to $t \sim 20$, corresponds to the timescale for the exponentiation rate to decline to $\sim 0.1$; 
cf.  Fig.~\ref{fig:exrate}). The diffusion limit in angular momentum becomes apparent 
only near the end of the runs. This is the case especially for smaller $N$, where the pre-diffusive systematic errors
reach large values before  diffusion set in. 
The post-exponential 
evolution of errors in the phase space variables
is characterised by a running power law index; for velocities, this is $\sim 2$ at $t \sim 20$, 
saturating towards smaller values (as errors reach order one) more slowly as $N$ increases.   
With these differences in error growth rates are associated differences in the $N$-scaling of the  'relaxation times' in the different variables, which we look at next.

\subsection{Relaxation times}
\label{sec:relax}

\begin{figure*}
\centering
					                    
\includegraphics[width=1.\columnwidth, trim = 0cm 0.cm 0cm 0cm, clip, angle = 0]{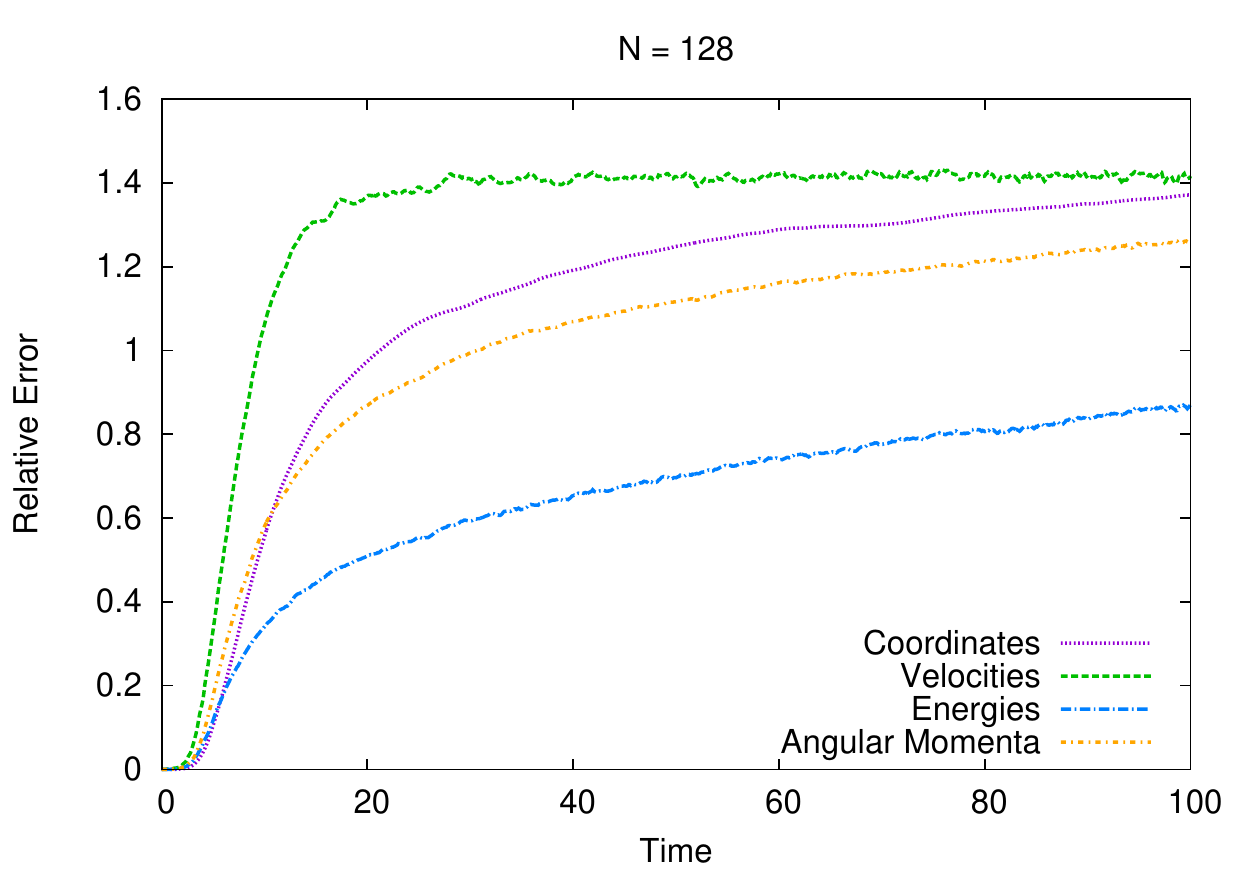}
\includegraphics[width=1.\columnwidth, trim = 0cm 0.cm 0cm 0cm, clip, angle =0]{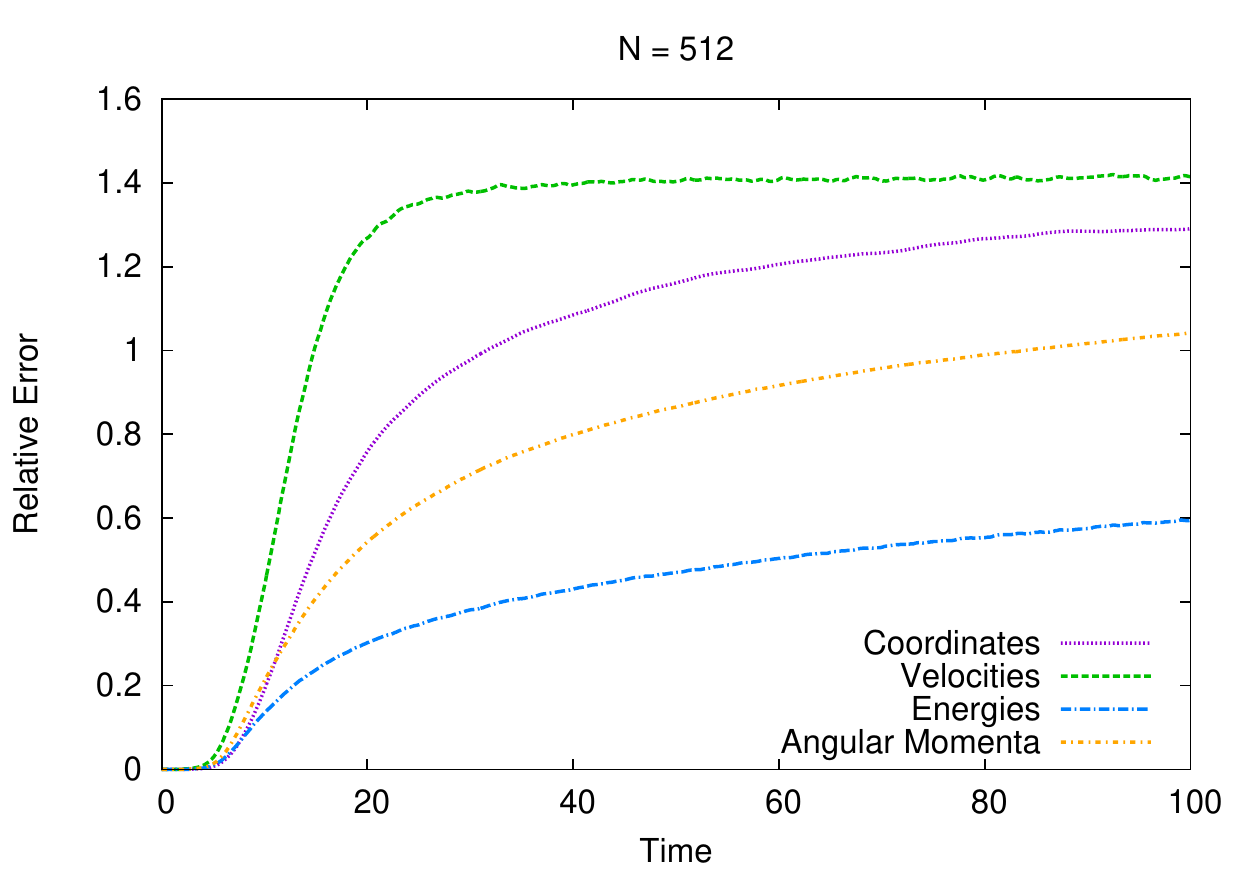}
\includegraphics[width=1.\columnwidth, trim = 0cm 0.cm 0cm 0cm, clip, angle = 0]{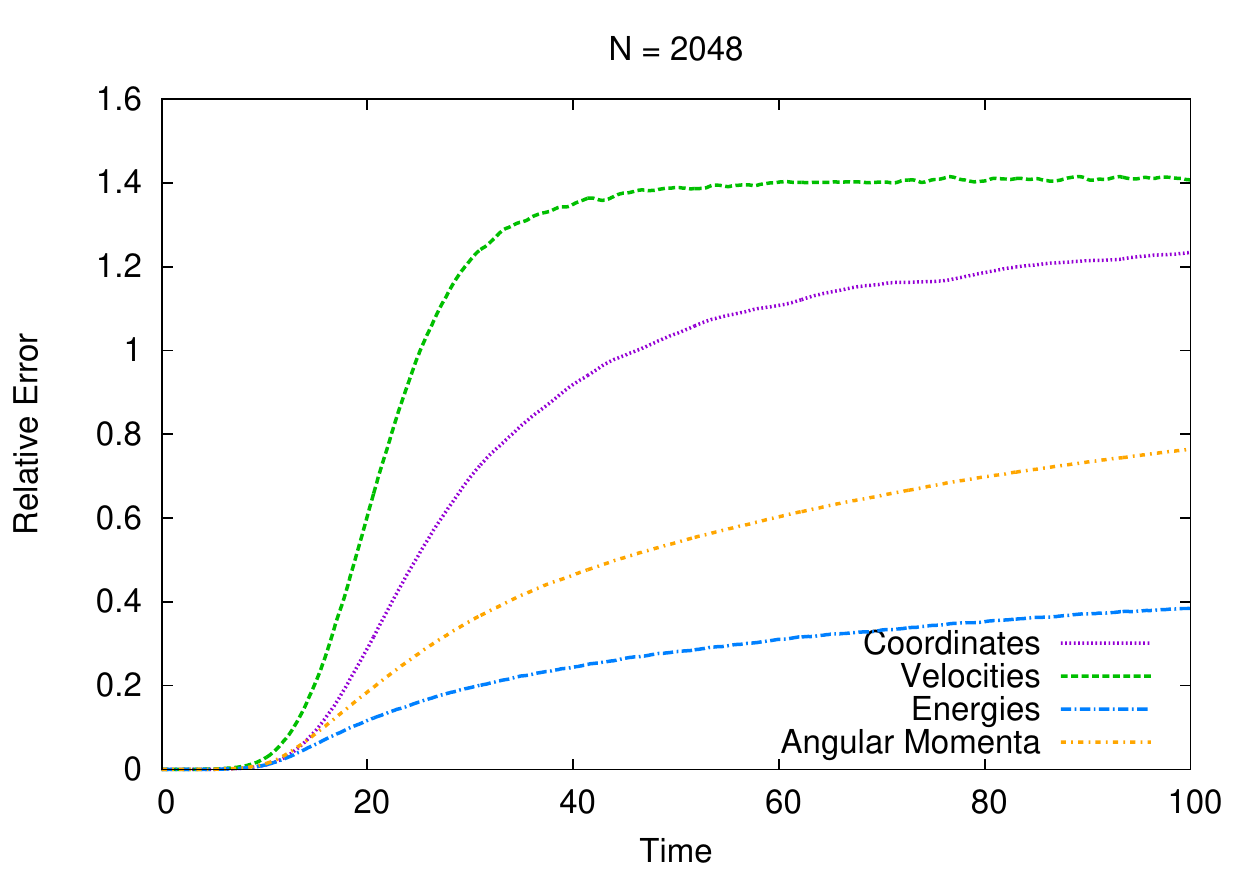}
\includegraphics[width=1.\columnwidth, trim = 0cm 0.cm 0cm 0cm, clip, angle =0]{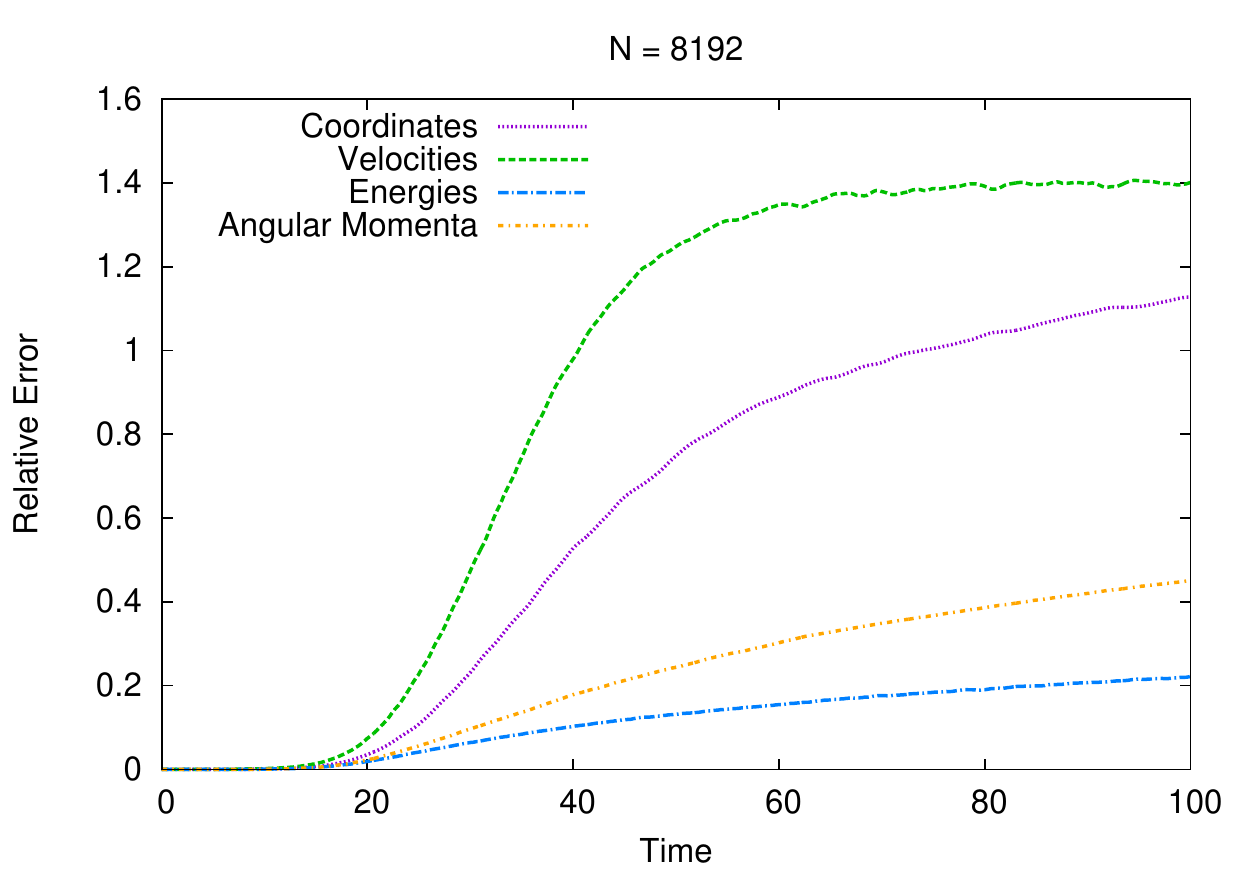}
\caption{Comparison of the error growth in the different variables. 
This is the same as in Fig.~\ref{fig:allvar} but now with fixed stepsize integrations and using a 
symplectic leapfrog algorithm. The errors are inferred by repeating the runs with the same initial conditions and 
halving the stepsize. For the largest $N$ runs ($N= 8192$) the results are similar to those in 
Fig.~\ref{fig:allvar}. For smaller $N$ some differences are apparent. In particular, angular momentum 
errors no longer initially follow that in the velocities for small $N$, and the diffusion limit 
is easily identified for all $N$.   As discussed in the text, an adaptive stepsize seems to limit the errors  
in the phase space variables in a way that does not translate into corresponding mitigation of errors 
in mean field conserved quantities. In particular, systematic error growth
in angular momentum seems to be enhanced at smaller $N$ when an adaptive stepsize is used.}

\label{fig:allvar_f}
\end{figure*}

\begin{figure}

\centering
					                    
	\includegraphics[width=1.\columnwidth, trim = 0cm 0.cm 0cm 0cm, clip, angle = 0]{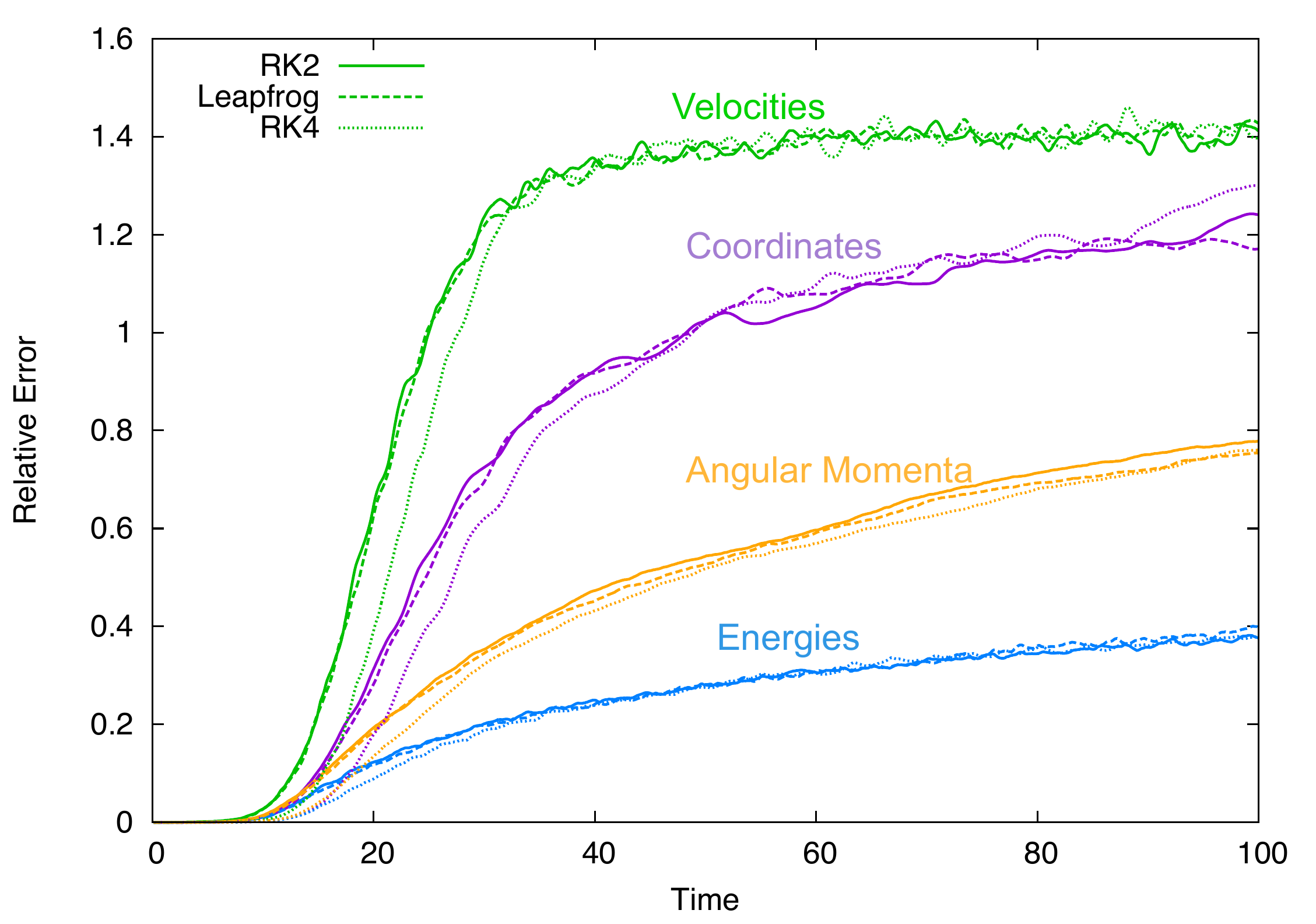}
\caption{Comparison of error growth in leapfrog, second order Runge-Kutta (RK2) and fourth order 
Runge-Kutta (RK4), with fixed stepsize. The errors are inferred by repeating the runs while 
halving the stepsize, with $N = 2048$.} 
    \label{fig:allvars_allmethods}
\end{figure}

We would  like to estimate  the timescale over which the error in a given variable $V$ reaches a threshold $h$
($V$ can either be a phase space variable or a mean field conserved quantity). 
During the  exponential phase, the growth in $V$ is $\delta V \sim e^{t/t_e} \delta V_0$, where $\delta V_0$ 
is determined by the value of the error at the start of exponential growth stage.  
The timescale to reach $h$ is thus $\ln \frac{h}{\delta V_0}$. 
According to the picture presented above (Section~\ref{sec:phase_space}),  
exponential growth persists until saturation time $t_c$, when 
$e^{t_c/t_e} \delta x_0 \sim R/\sqrt{N}$ (where as before $x$ is a spatial Cartesian coordinate).  
Assuming this same characteristic exponentiation rate for both $V$ and $x$, 
(as we already found), and that for $t \gg t_c$ the divergence can  be approximated by a power 
law in time, one may suppose that (for $t \ge t_c$) 
\begin{equation}
\delta V \approx \frac{R}{\sqrt{N}}   \Bigg(1+  \Bigg(\frac{t-t_c}{t_c} \Bigg)^s \Bigg)  \frac{\delta V_0}{\delta x_0}.
\label{eq:relax}
\end{equation}
 From Fig.~\ref{fig:logderis}, $1 \la s \la 2$  for $ t \ga 20$, before either 
diffusion ($s = 1/2$) or saturation ($s \rightarrow 0$) sets in (depending on the variable considered). 
For $t \gg t_c$,  one expects a  threshold $\delta V = h$ to be reached  on timescale
$t_{\rm relax} \sim t_c N^{1/2s}$. Thus, when a diffusion process is eventually dominant one expects, for a large enough threshold,  
an error growth that is roughly linear in $N$, akin to that deduced from two-body relaxation estimates. 
On the other hand,  for the systematic 
evolution --- characteristic of the error growth in the phase space variables before saturation sets in --- 
one expects the threshold to be reached on timescales scaling as $\sim N^{1/4}$ to  $\sim N^{1/2}$. 

Fig.~\ref{fig:Thresh} shows the `relaxation times', taken  to reach thresholds $h = 5 \%$ and $h = 33 \%$.  The timescales to reach 
the $5 \%$ threshold are dominated by the early exponential evolution.
The exponentiation rates do not vary much with $N$ or variable type. However, the 
error $\delta V_0$ at the start
of the exponential period however does, particularly due to the role of strong encounters as described in 
Sections~\ref{sec:velyev} and \ref{sec:comp_expo}  --- hence the 
dependence on $N$ and variable type. In the case of energy error growth,
which as was seen is relatively slow, the associated relaxation time at $h = 5 \%$ is large enough for the 
effect of the post-exponential evolution to be dominant for larger values of $N$. 
The growth of the timescale for the errors to reach $5 \%$ is in this case 
consistent with a scaling $\sim N^{1/4}$ for $N= 512$ and larger.  

For the threshold of $33 \%$, there is clear separation between the relaxation times associated with the different variables. 
For the energy,  a diffusion process is dominant at all $N$, and the relaxation time for error growth to reach $h = 33 \%$ is roughly linear. Indeed, it follows quite well the two-body relaxation form, with 
Coulomb logarithm factor $\gamma = 0.11$, as suggested by 
Giersz \& Heggie (1994).  (Note that the last two points, $N = 4096$ and $N= 8192$, are missing because the energy error 
does  not reach $33 \%$ over the timescales considered). 
For relatively small $N$ (up to $N = 512$),
the timescales associated with velocity and angular momentum errors are still dominated by 
the mechanisms of error growth that affect the angular momentum and velocity in similar manner 
(the pre-exponential, exponential, and systematic 
evolution driven by the multiplicative enhancement of encounters discussed in Section~\ref{sec:sim_mod}).  
For larger $N$, however, the timescale 
associated with the velocities grows as $\sim N^{1/4}$ (corresponding to $s \sim 2$ in~\ref{eq:relax}). 
The angular momentum relaxation to the $33 \%$ level, on the other hand, follows roughly a $N^{1/2}$
scaling. Except for the last point, when it steepens significantly 
as the diffusive contribution finally becomes dominant at $N = 8192$. 

\subsection{Propagation of errors in fixed stepsize integrations}
\label{sec:fixed}

The simulations presented in the previous subsections came with the advantage of an error estimate, calculated from a preset tolerance, which limits the maximal error even in the absence of softening. This is  the case where  
the central contradiction between the persistence of the exponential instability with non-saturating rate and the predictions of the collisionless limit is clearest. The local error, set by the tolerance,  
could be inserted into theoretical estimates of the dynamics of the growth of errors. It could thus be compared with the global error estimated using the reversibility criterion, which is linked to theoretical measures of dynamical information loss 
(as discussed in the Appendix). 

From a practical numerical point of view however, the method used in the previous subsections may seem detached from modern $N$-body simulations of collisionless systems. For these normally use symplectic methods, which are time symmetric and therefore reversible when the dynamical equations are.  The criterion of reversibility then becomes evidently irrelevant (except for tracking roundoff error if floating-point arithmetic is used). Simulations intended to model collisionless systems also invoke softening of the gravitational force. 

In this section, we repeat all our simulations while using a leapfrog integrator with fixed stepsize and softened force law. We run pairs of simulations, such that for every run there is an auxiliary one with the stepsize halved. 
An error estimate is derived by comparing the trajectories of the simulations with the larger stepsize with those with the smaller ones (started with exactly the same initial conditions).  
For our fiducial runs, we set the larger timestep at $\Delta t = 10^{-2}$ units and the (Plummer) softening  
to $\epsilon^2 = 0.001$.  For these values, the softening is small enough so as not to drastically 
suppress the dynamics of the exponential growth, and the RMS errors in particle energies are 
approximately the same as in the runs of the previous subsections. 
We have not undertaken a comprehensive examination of the effects of varying the softening and stepsize. 
However some trials --- in the range of $  10^{-4} \le  \Delta t \le 10^{-2}$  
and $  \times 10^{-5} \le \epsilon^2 \le  \times 10^{-3}$  --- suggest that the qualitative behaviour described below is generic. Though the absolute numerical values of error growth rates in each variable are sensitive to $\Delta  t$ and $\epsilon$, the relative growth rates in the different variables  
(coordinates, velocities, energies and momenta) are similar.  

Some of the results are shown in Fig.~\ref{fig:allvar_f}. For the largest $N$ runs ($N = 8192$), they are qualitatively the same as those in the corresponding Fig.~\ref{fig:allvar}. Except for the slower initial growth, due to the presence of softening, they are also quantitatively similar.  
Some differences are however apparent  between the smaller $N$ runs.  First, there is a steeper rise in the post-exponential error growth curves of the phase space variables, particularly the positions. In the light of the tests we have 
conducted (and describe further below), 
this is probably due to the lack of error control, provided in the previous  
subsections  by the preset tolerance. This leads to faster 
growth of errors in the phase space variables in the present runs, as compared to what was found before.  
If one assumes that encounters leading to growth of errors are local, and thus initially affect only
the velocities, then the larger errors in positions would reflect inter-encounter enhancement of errors initiated 
at discrete encounters.  
These would be mainly errors in phases of trajectories, which would explain the especially steep error
growth, in both phase space variables, when phase mixing is the dominant mechanism of growth, 
immediately prior to saturation. 
More importantly, Fig.~\ref{fig:allvar_f} also shows  slower
initial rise in the angular momentum errors, and a clearer transition to the diffusion limit 
than found in Fig. \ref{fig:allvar}, where an adaptive stepsize was used in conjunction with a preset tolerance.  
For smaller $N$, that prescription thus seems to {\it enhance} the error growth in angular momentum during the systematic growth stage, and to suppress its subsequent diffusive evolution. 

\begin{figure*}

\centering
					                    
	\includegraphics[width=1.\columnwidth, trim = 0cm 0.cm 0cm 0cm, clip, angle = 0]{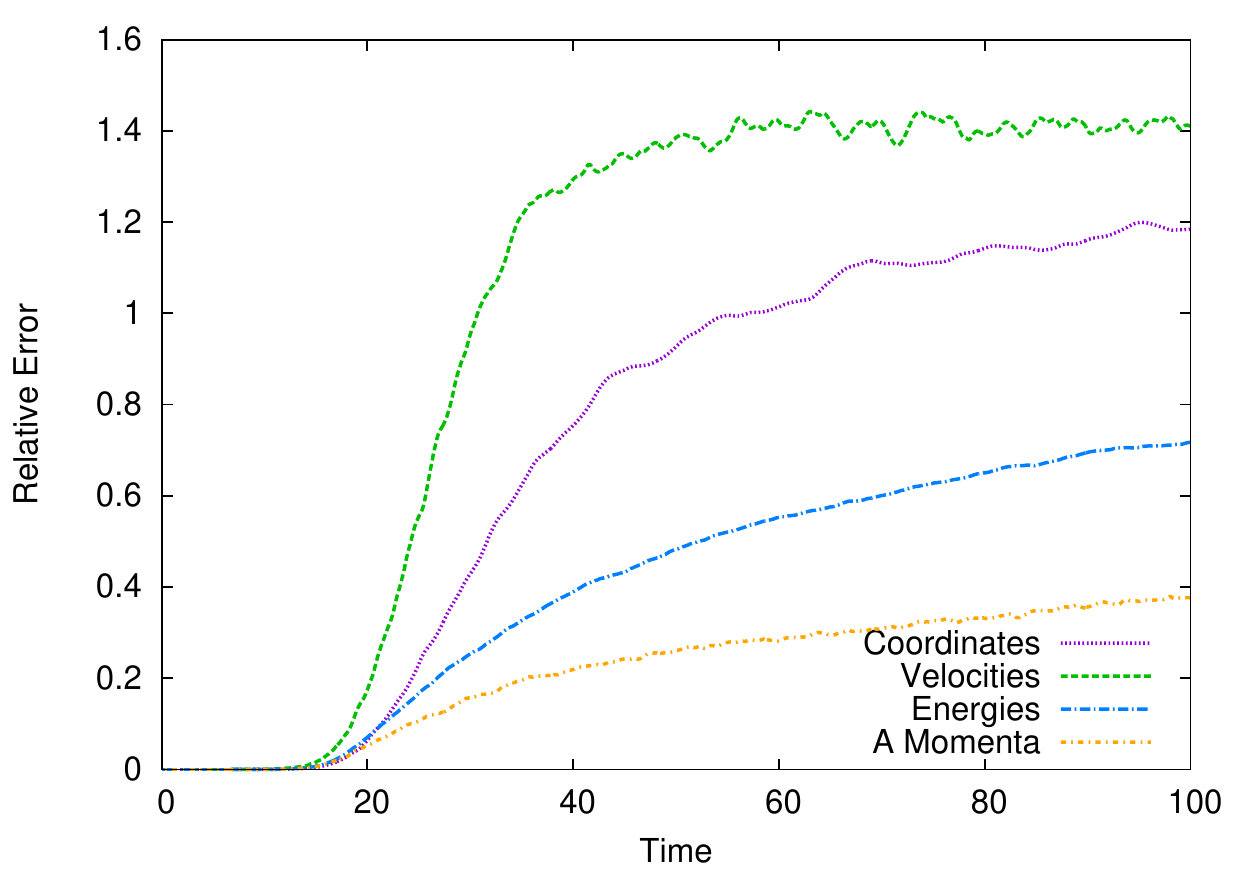}
	\includegraphics[width=1.\columnwidth, trim = 0cm 0.cm 0cm 0cm, clip, angle = 0]{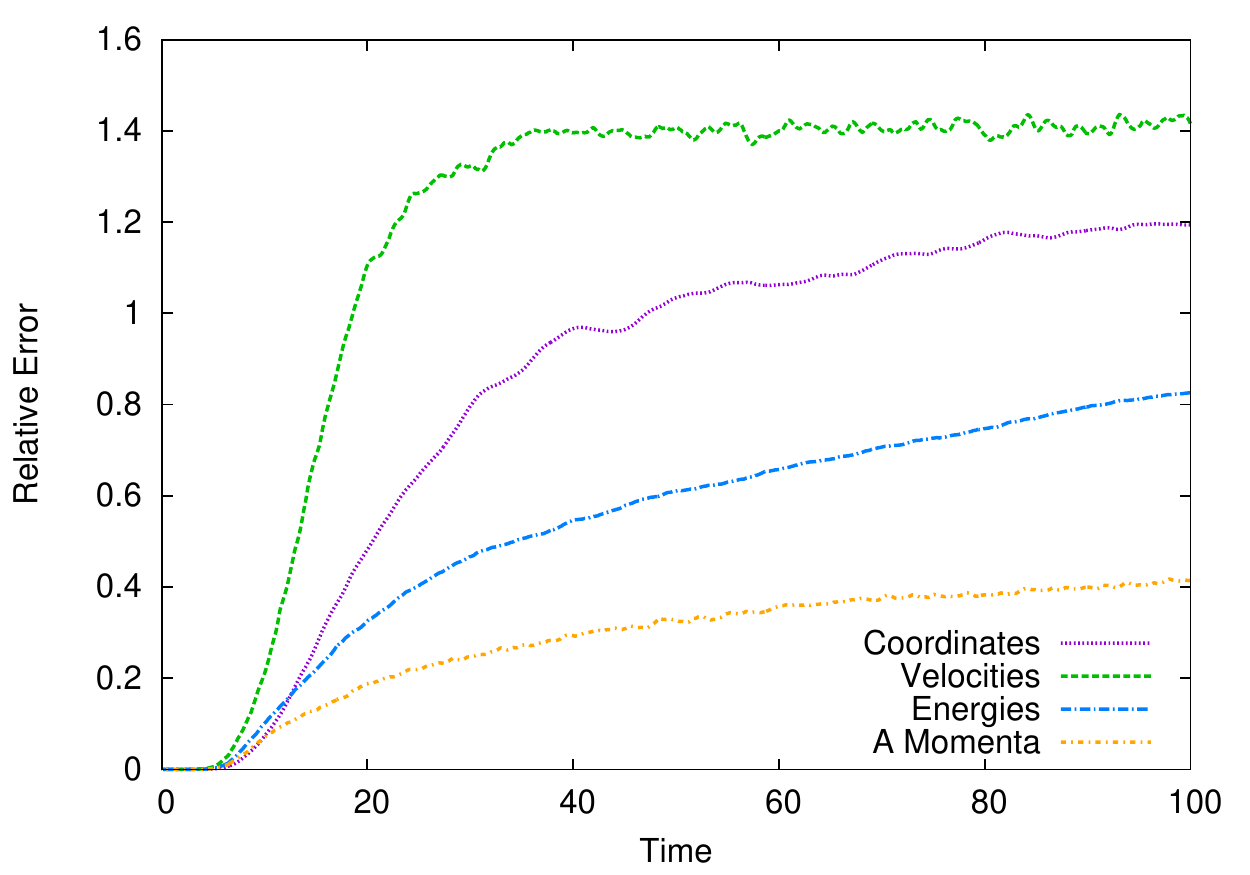}
\caption{Error growth evaluated using a fixed stepsize fourth order Runge-Kutta method and employing the reversibility criterion. With (Plummer) softening $\epsilon^2 = 0.001 $ (left panel) and  
$\epsilon^2 = 0.00025$ (right panel), and $N = 2048$.} 
    \label{fig:allvars_RK}
\end{figure*}

\begin{figure*}

\centering
					                    
	\includegraphics[width=1.\columnwidth, trim = 0cm 0.cm 0cm 0cm, clip, angle = 0]{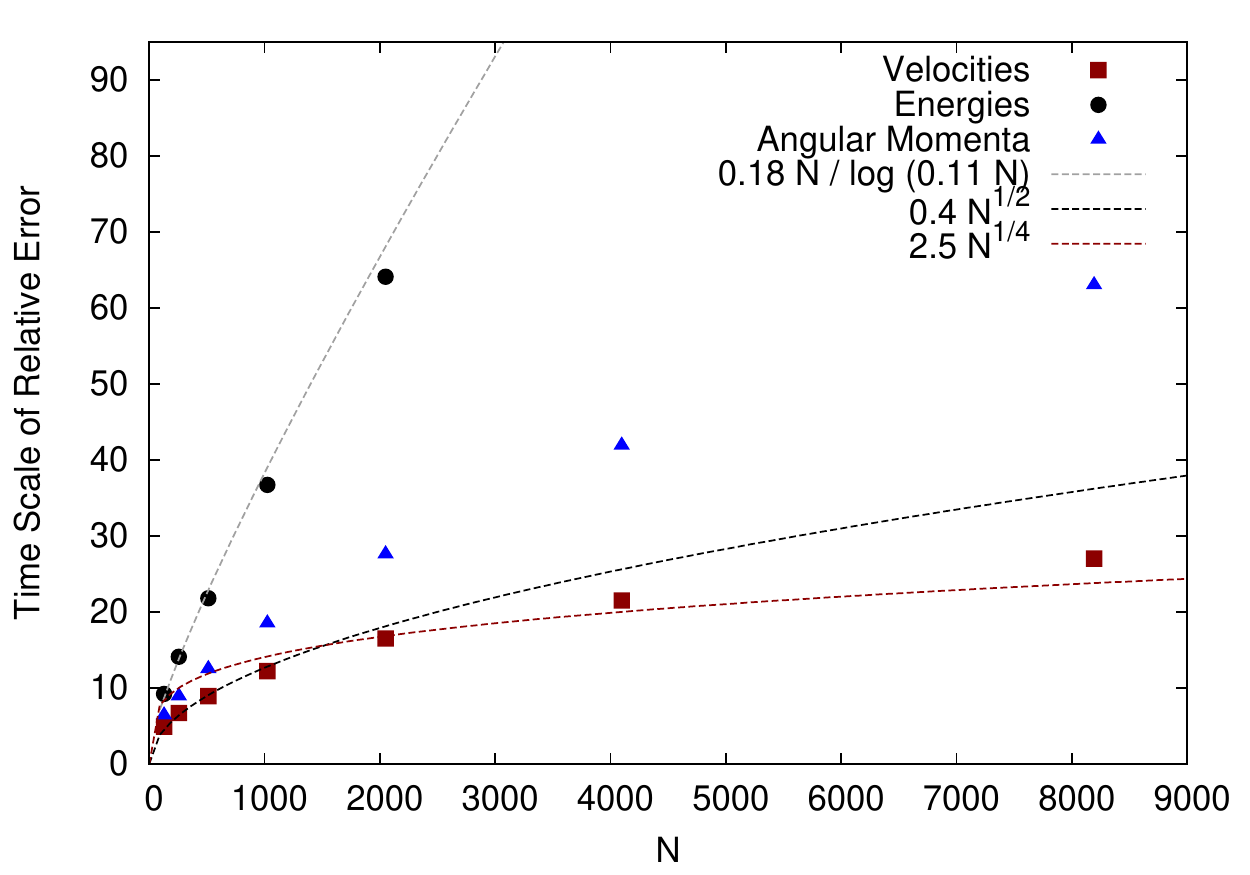}
	\includegraphics[width=1.\columnwidth, trim = 0cm 0.cm 0cm 0cm, clip, angle = 0]{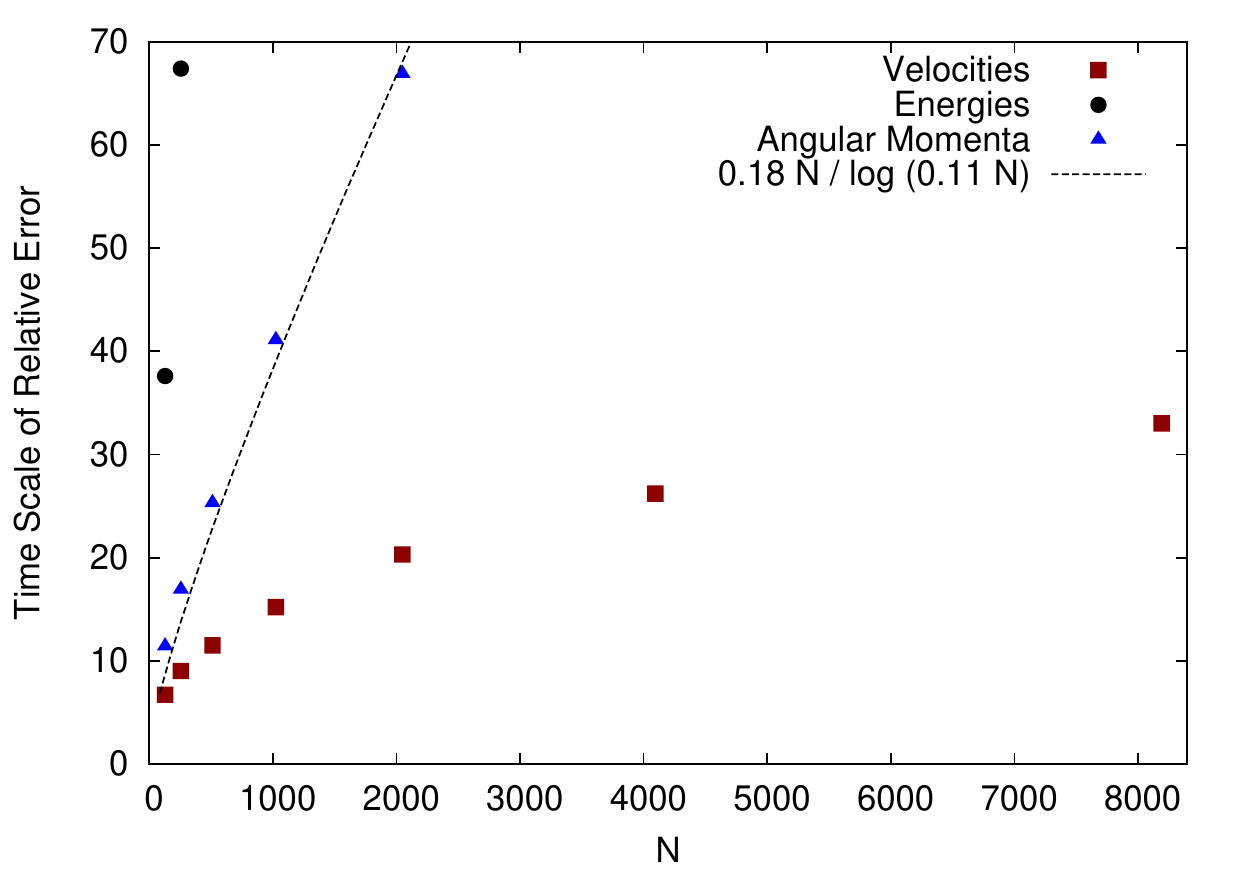}
						\caption{`Relaxation times', determined by evaluating 
differences between full and half timestep 
trajectories at the $33 \%$ (left panel) and $66 \%$ (right panel) threshold levels. 
At the $33 \%$ error level, the energy relaxation  can still approximately follows the two-body 
relaxation form with the same parameters 
as in Fig.~\ref{fig:Thresh}. The velocity error $N$-scaling is also similar at this threshold. 
The angular momentum relaxation $N$-scaling is however 
different.  Due to slower error growth, it no longer closely follows that of the velocities at smaller $N$, and the parametrisation used in  
Fig.~\ref{fig:Thresh} for larger $N$ (reproduced here for comparison),  no longer provides an adequate approximation.   
Nevertheless,  the diffusion limit, and associated 'two-body scaling', is only 
reached for the relatively high error threshold of $66 \%$.}
\label{fig:Thresh-2}
\end{figure*}

 In addition to checking that this general behaviour does not depend on  $\Delta t$ 
and $\epsilon$ --- at least in the range mentioned above --- we have checked that it persists when the smaller 
stepsize is reduced down to a tenth of the larger one (instead of half). 
We also  confirmed that it is not due to any structural 
evolution during the forward runs when the reversibility criterion is used. 
This was done by running the systems for 100 (and 200 and 300) 
units before halving the timestep and comparing the evolution with full and half stepsizes.

The differences are also not due to the use of the symplectic leapfrog integrator. This can be seen 
from Fig.~\ref{fig:allvars_allmethods}, where we present a comparison with what is obtained when a fixed stepsize second and fourth order Runge-Kutta are employed. The error growth associated with the second order Runge-Kutta is almost identical to the leapfrog, while the fourth order is somewhat smaller in magnitude, but still qualitatively similar for all variables. Thus, at least in terms of error propagation as measured here,   
it would appear that the advantage of the leapfrog method is confined to it being less expensive; needing one force evaluation per timestep instead of two for second order Runge-Kutta or four (for fourth order).

As a final test, we checked whether the discrepancies are related to the different error measures used;  
namely, comparing systems integrated 
with different  stepsize, as done in this section,  
as opposed to comparing the forward and reversed in time systems, 
as done previously.  
 In Fig.~\ref{fig:allvars_RK}
we present the results of a comparison run, 
using the reversibility criterion of the previous subsections but in conjunction
with a fourth 
order Runge-Kutta with fixed stepsize.  
Although the error growth is initially somewhat smaller in magnitude (compared to the lower left hand panel of Fig.~\ref{fig:allvar_f} and the RK4 curves in Fig.~\ref{fig:allvars_allmethods}), 
the qualitative behaviour is the same.   The quantitative differences can 
moreover be largely eliminated by varying the stepsize or the softening (as the right hand panel plot suggests).
The differences  with results of the previous subsections are thus not due to  
the different error measures used.

 It is therefore apparent that the slow error growth rate in momenta at smaller $N$, 
relative to other variables considered, is due to the use of a fixed stepsize, 
as opposed to one varied in accordance with a  
preset tolerance level as in the previous subsections.
 The energy errors are similar in the two cases (since, by construction, we could choose the  timestep and softening so as to ensure this). Since a preset tolerance,  
has limiting the maximum error in the phase space variables as criterion, 
it naturally limits the growth in these.
However, the control of errors in the phase space variables does not seem to reflect into a corresponding decrease in the angular momentum errors, at least in the systematic growth stage, when it is actually 
enhanced for smaller $N$. 
In  the diffusive stage, when it seems suppressed, the previous accumulation of errors is such that, when this stage is reached, the errors added are not large enough to become immediately apparent, except for $N= 8192$ or near the very end of the runs (as can be inferred for example  from Fig.~\ref{fig:logderis}).  In the fixed stepsize runs presented  here, on the other hand,  the pre-difussive error propagation is less steep and the diffusive growth is large 
enough to be measured as it sets in, even in the smaller $N$ runs.

As the angular momentum error growth in the simulations presented  here
displays clear diffusive evolution, like the energy,
we may expect an associated two body relaxation $N$-scaling, as previously found for the energy (Fig.~\ref{fig:Thresh}, right panel). 
Even if this should occur for a larger error threshold, as the errors in angular momentum 
are still clearly larger than those in energy. 
This is examined in Fig.~\ref{fig:Thresh-2}. 
For a $33 \%$ threshold, the energy errors still scale as in Fig.~\ref{fig:Thresh}.  For this error threshold, the velocities also scale approximately as before.
However,  the $N$-scaling of the angular momentum errors 
is steeper than in  
Fig.~\ref{fig:Thresh}. The parametrisations used there (and reproduced for comparison) is no longer adequate.
But the scaling is still slower than that characteristic of two body relaxation. 
For a $66 \%$ threshold, the velocity errors still vary relatively weakly with $N$, while the angular momentum errors now 
 follow the same two body relaxation form 
that the energy errors followed at the $33 \%$ threshold.  Thus, as in the case of energy, 
a clear transition from systematic to diffusive growth is finally detected. 
But the systematic stage of error growth still dominates till the errors are 
about twice those in the case of energy. The faster error growth in angular momentum implies 
that the timescales to reach that larger error threshold are similar for the energy errors to reach the $33 \%$ threshold. 
Assuming diffusive growth, a threshold larger by factor of two,  reached on the same timescale, 
translates to a `relaxation time' that is smaller by a factor of four.

\subsection{Summary and discussion}
\label{sec:sum_disc}

There are three qualitatively different ways through which errors in the trajectories of 
$N$-body systems can grow. They can grow systematically, as linear or polynomial functions in the 
number of time-steps $n$; or they can undergo a random walk and grow diffusively (as 
$n^{1/2}$); or they can grow exponentially.  Through examining the numerical reversibility of 
spherical $N$-body systems, we have observed three stages of error
growth incorporating these different aspects.  
Through the first two stages, the error growth is similar for the phase space variables 
(coordinates and velocities) and the mean field conserved quantities
(energies and angular momenta), while the third stage is characterized by
different  modes of  growth for the two types of variables. The difference becomes apparent  once
the `chaos' of the exponential instability has saturated, and the subsequent systematic growth, associated 
with the same sort of coherent scattering that lead to the exponential instability, has subsisted. 
The error in the phase space variables can then grow by phase mixing, while 
that of the mean field conserved quantities grows through a diffusion process. 

The initial stage occurs over timescales smaller than the exponential divergence timescale. 
The errors involved are principally truncation errors enhanced by strong encounters. 
This phase 
sets the initial errors that are inflated in the subsequent exponential stage.   The results of a  theoretical model
of the error growth through these two stages are shown in 
Fig.~\ref{fig:velyev}. The characteristic exponential timescale implied is a fraction of a dynamical time ( 
corresponding to  $0.15~t_D (r = a)$ and $0.08~t_D (R_v)$).  The error inferred from comparison of the forward and 
reverse dynamics is compatible to those estimated by the Runge-Kutta routine 
for relatively large $N$ but are much larger than these for smaller $N$. 
We interpret this in terms of the role of strong encounters, which diminishes as $N$ increases.

The exponential growth in coordinate errors does not saturate at dimensions comparable to the 
system size (as in systems where `chaos' can lead to global structural evolution); 
in fact, it already effectively ceases as the distance between two initially nearby trajectories 
becomes of order $1/\sqrt{N}$ the characteristic system size. 
We adapt a model due to GHH to interpret this 
phenomenon. In its context, the theoretical (dashed) lines of Fig.~\ref{fig:exrate} show that the coherent, 
multiplicative enhancement of errors 
due to encounters, that leads to exponential divergence, 
persists even after the saturation of the exponential growth. In fact 
up to when the effective exponential rate is 
$\la 10 \%$ its initial flat value.  
This implies that the systematic post-exponential evolution of the errors in the phase space variables 
is not immediately dominated by phase mixing, and neither is the post-exponential growth of errors 
in the mean field conserved quantities immediately diffusive. 
For angular momentum, diffusive growth in fact only becomes apparent at larger $N$ and late times 
when an adaptive timestep is used (cf. Fig.~\ref{fig:allvar}).
The growth of energy errors on the other hand always clearly displays diffusive behaviour, 
with growth rate $\sim t^{1/2}$, after about ten dynamical times. In contrast, the post-exponential evolution 
of the errors in the phase space variables can be locally characterised 
by a power law time evolution 
with index $ 1 \la s \la 2$.

The different modes of error growth lead to different estimates of the associated `relaxation times' ---
that is, the time the errors in a given variable  reach a certain fixed relative error level. For small thresholds (e.g., $5 \%$), the errors are dominated by the early growth. 
The relaxation times are very similar for the velocity and angular 
momentum, and also for the energy at lower $N$. At higher threshold (e.g. $33 \%$ 
level) the energy error growth is already fully diffusive, its $N$ variation is nearly linear and well explained 
by a two-body relaxation law with Coulomb logarithm  $\gamma = 0.11$.  
Assuming that error 
growth continues diffusively to the $100 \%$ level, 
the error growth time  to this level is  $t_{\rm relax}  = 0.26~\frac{N}{\ln (0.11 N)}~t_D~(R_v)$.  
This is quite similar to standard estimates of the two-body relaxation time.

As the velocity perturbations do not grow diffusively, the associated 
relaxation time has weaker $N$-scaling even at the $33 \%$ error threshold.   
And since angular momentum error growth initially follows that of the velocities, 
its relaxation time is also characterised by weak $N$-scaling. 
When a fixed stepsize is used, however,  
convergence of angular momentum errors to the diffusive limit is easier to identify.  
This is the case both when a symplectic (leapfrog) or fixed 
timestep Runge-Kutta methods are employed. In general, the results are almost identical 
(Fig.~\ref{fig:allvars_allmethods}).  Tests we have conducted however suggest 
that a preset tolerance level controls the 
errors in the phase space variables in a way that does not translate into a similar 
decrease in errors in the mean field conserved quantities, especially for smaller $N$.  
For the same energy errors, the angular momentum errors are generally smaller in 
the fixed stepsize runs, while those in the 
phase space variables are larger.

 Still, even in the case  of the fixed stepsize runs, the diffusion limit in angular momentum is reached 
at a larger threshold than in the case of energy. The corresponding relaxation time is about four times smaller.  This 
accelerated relaxation rate may  affect 
the structure of simulated objects,  particularly velocity anisotropies. It may 
also affect the spatial symmetry of 
triaxial systems, if the strong error growth in angular momentum extends to general action variables of regular orbits; in configurations  with a mixed  phase space of regular and chaotic orbits 
in the collisionless limit,
regular trajectories may diffuse into chaotic regions at a rate that is larger than that 
expected from standard two body relaxation estimates, thus accelerating morphological evolution.  
Such effects may be particularly important in 
small haloes and subhaloes, which are  identified in cosmological  simulations with relatively small number of particles, and far larger  errors in force calculations and time integration than in the simulations presented here.

 Even in the case of energy, 
slow diffusive error growth is only reached for relatively large error threshold 
(already at $ 33 \%$, in the simulations presented here). 
And the $N$-scaling of the errors can be far slower than the expected two body scaling until the diffusion threshold is reached. This  suggests that though there is convergence to the collisionless limit in a formal sense, which is important  from the theoretical point of view, as discussed below, this convergence is slow.

\section{Chaos and collisionless equilibria}
\label{sec:chao_coll}

In what sense are $N$-body gravitational systems `chaotic'?
Generally, one can identify chaos with the presence 
of a positive Lyapunov exponent. The exponential divergence 
between nearby trajectories does  imply the existence of at least 
one `finite time' positive exponent (strictly speaking, the exponents involve an infinite time 
limit, which is not well defined for systems with non-compact phase space as $N$-body 
systems).  Furthermore, for unsoftened systems, the associated timescale does not 
increase with $N$, and in that sense the instability does not saturate. For configurations with smoothed out potentials that 
are integrable, this directly contradicts the predictions of the collisionless limit, where all orbits are regular. 
Nevertheless, the results of the previous section confirm that this instability does not imply 
global structural evolution on the associated timescale, as it saturates on progressively smaller {\it scales} as $N$ increases.  
And mean field conserved quantities, such as energy and angular momentum, 
eventually evolve {\it via} a standard diffusive process.  
So in what formal sense are these systems chaotic; and does it matter, from a physical point of view?

A key issue concerns information loss, reflected in the 
numerical irreversibility and error propagation 
observed in the previous section, and whether it actually affects 
macroscopic parameters (spatial density, velocity dispersion etc.),
describing the gross statistical properties of the system (represented by 
moments of the one particle distribution function in case of 
collisionless systems).  In terms of numerical implications,  
if the local instability of trajectories 
affects the numerical image of systems whose physical properties should remain 
unaffected by the instability, then it degrades the faithfulness of simulations as well as their 
accuracy.  

To reconcile the fact of positive Lyapunov exponent with apparent lack of macroscopic 
evolution on the characteristic timescale, one should show that 
the associated loss of information decreases with $N$ at any macroscopic scale, and that it 
leaves the statistical quantities characterising macroscopic 
structure largely unaffected.  For this purpose, one can make use of the phenomenon of the saturation
of the exponential instability on progressively smaller scales, by relating
it to a measure of information loss on the phase space variables and conserved quantities. 
The concept of dynamical information loss can be made precise through the Kolmogorov entropy, 
which essentially measures the rate of increase in the information needed to follow a phase space trajectory in time
(e.g. Lichtenberg \& Lieberman 1992).  
A formal definition is technically involved, and a 
discussion as how it can be  applied to $N$-body systems, showing its close connection to effective irreversibility, 
is relegated to the Appendix.   Here we use a simpler definition based on the growth of coarse grained phase space volume.

Let $\Omega (0)$ represent a volume
in a six dimensional phase space of positions and velocities. 
Assume that $\Omega (0)$
is occupied by a continuous distribution of phase points, which define a set of initial conditions  
at a time $t_0$ (for finite $N$-systems, this implicitly assumes a statistical ensemble of such systems, 
similar to the average over simulations presented earlier, but as the number of ensemble members increases 
arbitrarily). If the  
 phase space is discretised into cells of volume $\Delta \Omega$, then  
 $\Omega (0)$ is obtained 
by counting the number of  cells that the phase points occupy. 
Even if a subsequent dynamical evolution of the phase flow obeys the collisionless Boltzmann equation
 --- and so 
Liouville's theorem applies on the 6-d phase space ---  
after $n$ timeteps, the volume  on the coarse grid will evolve such that 
$\Omega (n) \ge \Omega (0)$, as trajectories initially confined to a single cell 
at time $t_0$ can spread over many cells at later times.   In terms of the standard 
Boltzmann entropy, $\log \Omega (n)$, one can then define
\begin{equation}
K (\Delta \Omega) = \lim_{n \rightarrow \infty} \frac{1}{n} \log \frac{\Omega (n)}{\Omega (0)},
\label{eq:KS}
\end{equation}
which is non-vanishing if the volume increases exponentially.
The Kolmogorov entropy proper is the limit of 
 $K (\Delta \Omega)$ as  $\Delta \Omega \rightarrow 0$.~\footnote{A physically motivated discussion 
of this definition is given in Sagdeev, Usikov, \& Zaslavsky (1988).
A mathematical proof of its equivalence to the general definition, sketched in the Appendix, 
is given in Young (2003) for the
case of ergodic systems (where trajectories can come arbitrarily close 
to every point of the available phase space; e.g. BT).}
A nonzero value is associated with the presence of  positive 
Lyapunov exponents (and is directly related to their sum; 
e.g., Lichtenberg \& Lieberman 1992).
In this sense, due to the exponential divergence at infinitesimal scales, 
$N$-body systems have positive Kolmogorov entropy that does  not decrease with $N$, as the exponentiation 
rate does not.~\footnote{Note that, with infinite resolution, one can follow the fine grained evolution exactly, and the Boltzmann entropy would be constant if the collisionless Boltzmann equation applies (as $N \rightarrow \infty$). In the same way that with infinite information on the initial conditions one can, in principle, track  a trajectory  exactly, no matter how chaotic its evolution.  However,  as any infinitesimal perturbation will inflate as  $\delta x/ \delta x_0 \sim e^{k t}$, for arbitrarily small $\delta x_0$,  $k$ remains finite. In an analogous manner, there is a well defined limit in which 
$\Delta \Omega / \Delta \Omega_0 \sim e^{K t}$, and $K$ remains finite.}

Nevertheless,  the scale saturation of the exponential divergence  
implies that, at any finite volume 
resolution $\Delta \Omega$,  $K = 0$ as $N \rightarrow \infty$. 
For, given any two initially infinitesimally nearby trajectories and characteristic system size $R$, 
there is a maximal  
spatial distance $\sim R/\sqrt{N}$ (with corresponding maximal difference in velocities),
beyond which they cannot separate by means of the exponential divergence.
And if the progressively smaller scales over which the exponential instability occurs  
are not resolved, the exponential growth  
cannot be detected.  As the maximal volume in which the location of a  phase space 
point is uncertain at the conclusion of the exponential stage decreases (as $\sim 1/N^3$), so does  
the information loss suffered at the completion of  that stage. 
For an initial uncertainty  $\Delta x$ in phase space coordinates, it  goes as 
$\log \left(\frac{R}{\Delta x \sqrt{N}} \right)^6$. 
Eventually, in any numerical implementation, there is a maximal 
resolution imposed by machine precision. 
Thus,  in principle, beyond  $R/\sqrt{N} \sim 10^{-14}$ there would be 
no meaning for `chaos' due to the exponential instability, as far as double precision 
numerical implementation is concerned. 

The argument above is general. It concerns the loss of information 
due to the intrinsic exponential divergence of $N$-body systems and the scales on 
which it acts. This instability is invariably present at these scales, 
with the same timescale of a fraction 
of a dynamical time, regardless of the gross structure of the $N$-body configuration. 
After it saturates, however, the further growth of errors will depend on the 
detailed dynamics of the system at hand, for example on the spatial symmetry 
of the smoothed potential and whether it is time dependent or not. As 
the saturation occurs on smaller scales with larger $N$, this further growth of   
errors, which we now discuss,  becomes increasingly important.  
The arguments below pertain particularly to
configurations in dynamical equilibrium and which support only regular orbits 
in the collisionless limit, and are thus characterised 
by distribution functions that depend only on integrals of motion 
of these  orbits. This special case is the basis of much of 
classical galactic dynamics. In its context,  the loss of information is limited,  
quantitatively and qualitatively, by the slow post-exponential growth of errors, 
especially by the  transition to slow diffusive growth in the mean field conserved quantities
(that, is the integrals of motion in the smoothed potential).     

Suppose  that a set of trajectories (phase points) 
are started on an energy surface in phase space. As the system evolves, 
they will remain confined  
to volumes associated with a narrow energy range $\Delta E (t))$ that eventually grows only 
on the two body relaxation timescale.   
For spherical configurations with 
isotropic velocities, this is a shell with volume that can be expressed in terms of 
radial coordinates $r$ and $v$ in position and velocity space as 
\begin{equation}
\Omega (\Delta E) = \frac{\partial \Omega (E)}{\partial E} \Delta E = (4 \pi)^2 \Delta E \int_{0}^{r_{\rm max (E)}} v (E, r) r^2 dr,
\end{equation}
where $\Omega (E)$ is the phase space volume volume enclosed by energy surface $E$.
This scales as  $\sim \Delta E  v  r^3$, with $r$ and $v$ 
understood as typical radial phase space 
coordinates at energy $E$.
Next, consider initial conditions known to be homogeneously distributed within a cell of phase space of volume 
$\Delta \Omega (0) \sim  \Delta v_0^3 \Delta r_0^3$, where 
$\Delta r_0$ and $\Delta v_0$ are  uncertainties arising from coarse graining.  
Due to the loss of information associated with the dynamics, 
the phase point locations will eventually,  after a time $t_E$ say, be uncertain. 
As we saw in the previous section, the error growth rate in energy
is much smaller than that in angular momentum. One 
may then consider an intermediate state when 
the phase points  would be uncertain only within
the energy shell $\Delta E$, even after loss of information on other mean field 
conserved quantities is essentially complete.  
In this case, the location of the phase points will be uncertain within a volume  
$\Delta \Omega (t_E) = \Omega (\Delta E)$, which is  $\sim \Delta E  v  r^3$.  
If initially $\Delta E_0 \sim (\Delta v_0)^2$,   
then assuming strict energy conservation ($\Delta E = \Delta E_0$),
but complete loss 
of information on the location of phase space points within the energy 
shell, implies
\begin{equation}   
\frac{\Delta \Omega (t_E)}{\Delta \Omega (0)} 
\sim \frac{v r^3}{\Delta v_0 (\Delta r_0)^3}.  
\end{equation}
If  $\Delta v_0/v = \Delta r_0/r \sim 10^{-14}$, say, then
$\Delta \Omega (t_E)/\Delta  \Omega (0)   \sim 10^{56}$.    
Thus, the exponential instability, which seeds the process of loss of information 
over the energy shell, does have an  effect in this sense.  
However, although the volume multiplication is large, the timescale over which it occurs increases with $N$, 
since for large enough $N$ most error accumulation occurs after the saturation of the exponential instability.  
According to estimates of Section~\ref{sec:relax},
the timescale associated with the error growth in the phase 
space variables should then scale as $N^{1/4} - N^{1/2}$.
Furthermore, even with total uncertainty on the phase space variables within the energy shell, 
the loss of information 
is still quite small compared to that accompanying  an energy uncertainty also of order one, 
in which case an additional factor of $10^{28}$ in volume ratio is involved. 
And this last process  occurs over the still longer timescale of 
diffusive error growth, scaling as $\sim N$.

More importantly, information loss that occurs while the integrals of motion 
remain well conserved
primarily reflects uncertainty in the phases of trajectories. It is in this 
sense dynamically trivial, and its implication concerning simulated systems relates to the 
accuracy of the results and much less to the faithfulness of the representation.
For example, suppose the system (as is the case of the systems considered in the previous section), 
can be characterised by a distribution function  $f = f(E)$.
As long as energies are reasonably well conserved along trajectories,  
even total loss of information on the phase space coordinates
does not imply  corresponding evolution in the 
distribution function. Hence it should not significantly affect the macroscopic state 
 of the self gravitating 
configuration, described by $f$ and its moments. 
Indeed, if phase points were labelled only by their energies, and 
the phase space coarse-graining done solely in terms of energy  shells $\Delta E$, 
then the coarse grained $\Omega (\Delta E)$ would not change dramatically over timescales 
much smaller than the energy relaxation time.  The entropy given by equation (\ref{eq:KS}) 
 would still be non-zero over infinitesimal scales, as on these scales the exponential divergence affects the
mean field conserved quantities in the same manner as the phase space variables. But after it saturates, 
the information loss is limited, as the slow diffusive stage of error growth is reached. 
A similar situation would arise if the distribution function 
depended also on momenta, $f = f(E, L)$, or was a function of action 
variables in the case of general integrable potentials.

To sum up. For any finite phase space resolution, the Kolmogorov entropy tends to zero as 
$N \rightarrow \infty$. 
In particular, for any spatial resolution $\ga R/\sqrt{N}$ it tends to zero.
The information loss incurred during the exponential stage is thus
increasingly limited with increasing $N$, as its saturation scale shrinks.
From this perspective, the exponential 
instability in large $N$-body systems does not imply that they undergo significant 'chaotic mixing', which  
explains why they do not generally display relaxation and evolution of macroscopic quantities on the exponential 
timescale, a characteristic of systems where the exponential instability saturates on 
larger scales (Kandrup \& Sideris 2003).    
Second, 
although dynamical 
evolution leads to large loss in information regarding the phase space coordinates of trajectories, 
beyond the exponential regime this occurs on timescales that increase with $N$. 
The loss is moreover small compared to that accompanying the error growth in the mean 
field conserved quantities, 
which takes place on a still longer time scale with steeper $N$-scaling, akin to that of standard two 
body relaxation. Thus, for  
systems characterised by a distribution function that depends on the mean field conserved quantities, the 
the gross structure is largely unaffected on smaller timescales, despite the large loss 
of information on the phase space coordinates of trajectories. 
 
This gives confidence in the convergence towards the collisionless limit and its predictions,   
despite the persistence  of the exponential instability on infinitesimal scales.
It  gives credibility to the practice of modelling gravitational systems through orbital integration in smooth 
potentials --- the process of mapping characteristics curves of the collisionless Boltzmann equation ---
and credence in the faithfulness of $N$-body simulations, despite the loss of information inherent in numerical integration.
It should be noted nevertheless that the slow scale 
saturation of the exponential stage --- which is followed by systematic growth of errors before 
slow diffusion sets in, even in the case of the mean field conserved quantities --- points to the slowness 
of the convergence. Indeed, as we saw in the previous section, the diffusion limit, 
and associated steep $N$ scaling for error growth, 
was only reached at the $33 \%$ error threshold in the case of energy. 
This threshold was even larger in the case of  angular momentum 
(indeed, in this case,  the diffusion limit was clearly reached at all, for all $N$, 
only in the fixed timestep runs).

\section{Conclusion}

We have  examined two related aspects of $N$-body gravitational systems; 
the dynamical divergence of initially nearby solutions and the faithfulness of their numerical simulations. 
We did this by considering  the simple case of  spherical systems in dynamical
equilibrium, running a suite of small $N$  but high accuracy simulations, in the range 
of $N = 128$ to $N = 8196$, while quantifying the direct consequences of the growth of errors. 
 This was done by studying the numerical reversibility of solutions
advanced with a preset error tolerance 
(set to $10^{-8}$), and also by comparing fixed timestep integrations 
conducted with different stepsizes. 
The latter method mainly employed the widely used symplectic leapfrog integrator, with 
some comparison runs using second and fourth order
Runge-Kutta. The force law was softened, as opposed to the case in runs with adaptive stepsize, where 
there was no softening.

Three phases of error growth can be distinguished. 
A simple calculation and numerical results show that the errors add up randomly  
before the well known local exponential instability 
of $N$-body systems sets in.  As have been found in other studies, the exponential growth time is virtually 
independent of $N$. However the pre-exponential growth is sensitive to $N$; as measured by the irreversibility of the 
computed dynamics, it is larger than the estimated truncation error for relatively small $N$ but significantly  smaller 
than that error for larger $N$.  Moreover, the exponential instability saturates at progressively smaller spatial scales,  of order $1/\sqrt{N}$ the system size, as $N$ increases. 
A simple model, originally due to Goodman, Heggie \& Hut (1993), was adapted to study this phenomenon quantitatively.  

These first two phases of the error growth  
set the stage for the third phase, by providing its initial conditions.
This stage starts with systematic growth, driven by the same mechanism
of coherent, multiplicatively reinforcing enhancements, born of encounters of the same sort 
that lead to the exponential instability. Subsequent to this, the growth of errors is 
qualitatively different  in the case of the phase space variables (coordinates and velocities) and the 
conserved quantities of the mean
field dynamics (energies and angular momenta); 
the former grow through phase mixing, while the latter grow according to a diffusion ($\sim t^{1/2}$) law. 
This implies different timescales for error growth and their variation with $N$.

The convergence towards the diffusion limit is always clear in the case 
of energy, where  the inferred `relaxation time'  of error growth
is of the form of that obtained from 
standard two-body relaxation theory.  The convergence towards that limit is harder to identify 
in the case of angular momenta, especially when a variable timestep is used. In this case, for a similar 
level of energy errors, the  preset tolerance level limits the error growth in the 
phase space variables in a way that does not translate into similar control 
in angular momentum errors, especially for smaller $N$
(for larger $N$, there is general agreement with the results of the fixed stepsize runs).  
Even in the case of the fixed stepsize runs however, the diffusion limit in
angular momentum is reached only at an error threshold about double that of energy 
(and already when the relative error has reached $66 \%$). 

The relatively short angular momentum error growth times may have  
consequences for estimating the effect of discreteness effects in simulations.  
Velocity anisotropies may be particularly affected by the larger growth rate of angular momentum errors, 
and  if the phenomenon applies to general action variables in non-spherical potentials, associated errors can affect the shapes of the corresponding objects (presumably rendering them more spherical). 
A case in point concerns finite-$N$ configurations with mixed phase space, 
supporting  chaotic as well as regular orbits in the collisionless limit. Here, regular trajectories may diffuse 
into chaotic regions on timescales shorter than that expected from standard two body relaxation 
estimates. Once in such regions, the action variables may not be even approximately conserved  or defined, 
and the trajectories may fill the available chaotic region of phase space on a smaller timescales still. 
Other situations of interest include those involving time dependence of the potential.
When this is slow, action variables become adiabatic invariants. The level 
to which they may be conserved in finite-$N$ systems may also be affected by 
relatively rapid error growth rates.   
Such situations, as well more general ones (for example when mergers are involved), 
may be worth examining in detail in future studies.

The exponential instability persists, at the infinitesimal level,  with a rate that does not 
decrease with $N$. In this sense, arbitrarily large-$N$ unsoftened gravitational systems are `chaotic'
even for configurations that support only regular orbits in the collisionless limit.  
However, when the effect of the local instability is viewed in terms of information loss 
on a finite resolution scale, a different picture emerges.  
Because the exponential divergence saturates at increasingly smaller scales, the 
Kolmogorov entropy, which must be positive for a chaotic system,  tends to zero at 
at any non-infinitesimal resolution level as $N \rightarrow \infty$. This is because,
following the completion of the exponential divergence,
 the volume
in which a phase space trajectory can be localised, 
stage decreases as $\sim 1/N^3$. A finite resolution 
level  is inherent in any practical representation, which is ultimately limited by machine 
precision (or by softening, which, when large enough, 
has been shown to increase the exponential timescale as $N$ increases: 
e.g. GHH; El-Zant 2002). 
This necessarily limits the sense in which $N$-body systems can be termed 
`chaotic'.   Indeed, the loss of information is especially minimal if it is viewed in terms of mean 
field conserved quantities. As long as these are reasonably well conserved  
a distribution function depending on integrals of motion is also correspondingly conserved, and 
the dynamics  tends to its mean field counterpart,
despite the persistence of the exponential instability on infinitesimal scales.

Thus, despite the persistence of the exponential instability --- which 
for spherical systems  in equilibrium is in apparent, direct contraction to what is expected in the collisionless limit --- 
our results support the standard assumptions of galactic dynamics, based on that limit. They give credence to the associated practice of modelling  large-$N$ systems through orbital integration in smooth 
potentials, which is perhaps of increasing importance for understanding of Galactic dynamics 
in the age of GAIA, and  confidence in the faithfulness of $N$-body simulations.     
Nevertheless, the weak  $N$-scaling of the error saturation level of the 
exponential instability entails a  similarly weak scaling ($\sim N^{1/4}$ to  $\sim N^{1/2}$) for the 
propagation of errors during the subsequent stage of systematic growth. Even in the 
case of the mean field conserved quantities, this  dominates before the diffusion limit, 
with its steeper scaling  ($\sim N$), is reached. And since this is reached at  relatively 
large error thresholds,  the convergence  to the predictions of the collisionless limit to high precision 
(at the $10 \%$ error levels say) is slow, even when measured in terms of energy errors.    
The errors in the phase space variables never reach a diffusive stage; they continue to propagate
systematically through phase mixing, and retain a weak $N$ scaling for all error thresholds.

\section*{Acknowledgements}

We would like to thank the referee for  careful readings of the 
manuscript and comments that lead to marked improvement. 
This project was supported financially by the Science and Technology Development Fund (STDF), Egypt. Grant 
No. 25859.





\appendix

\section{Kolmogorov entropy and effective ireversibility}

Assume we are dealing with timescales that are small compared to the energy relaxation time, 
and that the system is described by a 6-d probability phase space distribution function $f = f (E)$
obeying the collisionless Boltzmann equation (and so $f$ is constant along particle trajectories). 
Discretise the phase space into cuboidal cells, each with volume $\Delta \Omega$, and follow the evolution 
of a particle trajectory from specified initial conditions. Then record the cell number 
$c_i$ at which  a  trajectory is located at time times $t_i$, with $i = 1,..,n$.  
The Shannon information 
entropy of this partition of cells under the action of the trajectory is  
\begin{equation}
S = - \sum_i^n p_i \log p_i,
\end{equation}
where probabilities $p_i$ are proportional to the time the trajectory spends in cell $i$.

Now consider a group of trajectories initially localised in one of the phase space  cells. 
The Shanon entropy for this system of trajectories generally increases as the system evolves, 
and is maximized as the probabilities $p_i$ 
equate, so that $p_i \rightarrow f(E)$ on all cells. 
The Kolmogorov entropy distinguishes between any general increase in Shanon entropy 
(e.g. due to phase mixing) and that associated with 
exponential instability. For our purposes the concept can be developed as follows.

Suppose that after at time $t_1$ we take a distribution $f$ of points over the cuboidal phase space cell 
$c_1$ where our integrated trajectory resides. If we reverse back the trajectories in this cell to time $t_0$, the collisionless evolution
ensures that the fine grained 
density $f$ is conserved and the distribution will end up under the reversed dynamics in cell $c_{0,1}$ of same volume as $c_1$. 
If there is no divergence between neighbouring trajectories  the shape of the cell is conserved during the evolution, and volume 
conservation implies that cell $c_{0,1}$ will coincide with $c_0$ where they all started. 
In the presence of divergence however, trajectories reversed from $c_1$ may fall in cells other than
$c_0$. In fact the cell $c_{0,1}$ containing the reversed trajectories will be distorted by the evolution and will in general not be 
cuboidal anymore. The intersection of $c_{0,1}$ with $c_0$ will generally define a cell of smaller volume. 
The continuation of this procedure for times $t_2, t_3..., t_n$ defines progressively smaller cells, determined by the 
intersections of $c_{0,1}, c_{0,2}, ..., c_{0,n}$ with $c_0$.  Now if we label the cells defined by these intersections as $d_j$ 
(with $j=1,.., n$),  and 
repeat the process for different trajectories of the system (with initial conditions sampled from the distribution function $f(E)$),
we can define a finer partition of the phase space.  The associated probabilities $p_{d_j}$ 
(which are proportional to $f(E)$ integrated over smaller and smaller volumes of the intersection cells), 
decrease with increasing maximal timestep $n$.  

The Kolmogorov entropy  measures
the increase per time-step of the Shannon information entropy of this increasingly finer partition as volume $\Delta  \Omega$ and the time-step 
$\Delta t$ both go to zero. This corresponds to the loss of information per time-step, which leads in practice to the inability to numerically 
track and reverse a trajectory.  The reason we obtain finer partitions from the reversed dynamics --- and corresponding increase in Shannon entropy ---  was that trajectories that were initially confined to a cell $c_i$ at time $t_i$ could be dynamically evolved into occupying  many cells (hence the increase in coarse grained volume). 
This formally relates the phenomenon of effective irreversibility, 
due to information loss, to the local exponential divergence between nearby trajectories. 
It also relates the definition here to the more intuitive one given in the body of the paper.



\bsp	
\label{lastpage}
\end{document}